\documentclass[11pt]{article}

\usepackage{cancel}

\newcommand{\be}{\begin{equation}}
\newcommand{\ee}{\end{equation}}
\newcommand{\bea}{\begin{eqnarray}}
\newcommand{\eea}{\end{eqnarray}}
\newcommand{\bg}{\begin{gather}}

\newcommand{\bseq}{\begin{subequations}}
\newcommand{\eseq}{\end{subequations}}

\renewcommand{\ln}{\mathop{\rm ln}\nolimits}

\renewcommand{\Im}{\mathop{\rm Im}\nolimits}
\renewcommand{\Re}{\mathop{\rm Re}\nolimits}

\newcommand{\lambdabar}{\lambda\mkern-9mu\raisebox{0.15ex}{\rule{0.45em}{0.08ex}}}

\def\be{\begin{eqnarray}}
\def\ee{\end{eqnarray}}

\usepackage{revsymb}
\usepackage{amsmath,amssymb}
\usepackage{array}
\usepackage{graphicx}
\usepackage{graphicx}
\usepackage{wrapfig}
\usepackage{color}
\usepackage{tensor}
\usepackage{hyperref}
\usepackage[sort,compress]{cite}
\usepackage{tikz}
\usepackage[T1]{fontenc}
\usepackage{multirow}
\usepackage{longtable}
\usetikzlibrary{decorations.pathmorphing, decorations.pathreplacing, decorations.shapes} 

\numberwithin{equation}{section}

\begin{document}
\title{\textbf{
Tidal Love numbers of wormholes 
as black-hole mimickers}}
\vspace{1cm}
\author{ \textbf{ Sergey N. Solodukhin   and   Vagif Tagiev }} 

\date{}
\maketitle
\begin{center}
    \emph{Institut Denis Poisson UMR-CNRS 7013,
  Universit\'e de Tours,}\\
  \emph{Parc de Grandmont, 37200 Tours, France} 

\end{center}

\vspace{0.5cm}





\begin{abstract}
We study the dynamical scalar tidal Love numbers of wormholes that provide viable mimickers for black holes, focusing on thin-shell Schwarzschild and Damour-Solodukhin geometries. Using a matched near- and far-zone expansion, we determine their tidal response in the low-frequency regime. The presence of a long throat introduces an additional characteristic scale and naturally separates the modes into two classes. Super-throat modes probe the global wormhole geometry and are sensitive to both asymptotic regions, whereas sub-throat modes probe only one side of the wormhole and effectively perceive the throat as a black-hole horizon. We derive the scalar tidal Love numbers analytically for both classes of modes and show that their dissipative parts exhibit distinct low-frequency behavior, reflecting whether one or both potential barriers participate in the scattering process. We further find that, as the wormhole approaches the black-hole limit, the super-throat contribution becomes progressively negligible, while the sub-throat response smoothly reduces to that of a Schwarzschild black hole. These results demonstrate that the tidal response of wormholes depends crucially on whether the perturbation probes the global structure of the throat.
\end{abstract}
\rule{7.7 cm}{.5 pt}\\
\noindent ~~~ {\footnotesize  sergey.solodukhin@univ-tours.fr\\
                              vagif.tagiev@univ-tours.fr }

\newpage
\tableofcontents
\pagebreak
\section{Introduction}\label{sec:introduction}

The first direct detection of gravitational waves \cite{LIGOScientific:2016sjg, LIGOScientific:2016aoc}, followed by the rapid development of gravitational-wave observations \cite{KAGRA:2021vkt}, has opened a new window onto the study of compact objects in our Universe. Gravitational-wave astronomy has rapidly become a powerful tool for probing neutron stars, black holes, and compact objects more generally. Gravitational waves emitted by binary systems provide valuable information about the nature of their sources while simultaneously offering a means of testing general relativity (GR) in the strong-field regime \cite{LIGOScientific:2020tif}.

 The coalescence of a compact binary can be broadly divided into three stages. The first is the inspiral phase, which is well described within the post-Newtonian (PN) framework \cite{Blanchet:2013haa, Buonanno:1998gg}. This is followed by the merger phase, a highly relativistic regime dominated by strong-field effects that generally requires numerical-relativity calculations \cite{Blanchet:2009sd, Campanelli:2005dd}. The final stage is the ringdown, during which the merger remnant relaxes toward equilibrium through the emission of gravitational waves. The ringdown is governed by quasinormal modes (QNMs) \cite{Vishveshwara:1970zz, Nollert:1999ji, Dreyer:2003bv, Berti:2009kk, Konoplya:2011qq}, whose spectrum encodes information about the nature and geometry of the compact object.

Throughout this evolution, the binary emits gravitational waves that carry away energy and angular momentum, causing the orbital separation to decrease. The amplitude and phase of the gravitational-wave signal therefore encode information about both the orbital dynamics and the intrinsic properties of the compact objects. Among these properties, tidal deformability plays an important role during the inspiral. When the two bodies are widely separated, they can be approximated as point particles moving along their respective orbits. As the separation decreases, however, each object is subjected to the tidal field of its companion, leading to a deformation of its internal structure. These tidal deformations modify the orbital dynamics and leave characteristic imprints on the emitted gravitational-wave signal \cite{Flanagan:2007ix, Hinderer:2007mb}. The tidal contribution to the inspiral waveform is encoded primarily in its phase, with the leading-order tidal effects entering at 5PN relative to the point-particle contribution.

The quantities that characterize the tidal response of a compact object are known as tidal Love numbers (TLNs).\footnote{The concept of Love numbers was originally introduced by A. E. H. Love to describe the tidal deformation of the Earth \cite{Love:1909}.} These quantities are sensitive to the internal structure of the object. For neutron stars, for example, the tidal response depends strongly on the equation of state of nuclear matter \cite{Hinderer:2009ca}.

Remarkably, in four-dimensional GR, the leading contribution to the black-hole tidal response—the static Love numbers—vanishes for both nonrotating \cite{Fang:2005qq, Binnington:2009bb, Damour:2009vw, Kol:2011vg, Hui:2020xxx} and rotating black holes \cite{Poisson:2014gka, LeTiec:2020spy, LeTiec:2020bos, Charalambous:2021mea}. In other words, black holes exhibit no static deformation in response to an external tidal field.
This vanishing of the static Love numbers has been related to a possible ``hidden symmetry'' of four-dimensional black holes \cite{Charalambous:2021kcz, Charalambous:2022rre, Sharma:2024hlz, Ghosh:2026vig}. Such behavior is rather special: it does not generally occur for other compact objects in GR \cite{Cardoso:2017cfl, Mendes:2016vdr, Pani:2015tga, Chakraborty:2023zed}, for black holes in modified theories of gravity \cite{Cardoso:2018ptl, Barbosa:2025uau, Cano:2025zyk}, for black holes in higher dimensions \cite{Chakravarti:2018vlt, Pereniguez:2021xcj, Charalambous:2023jgq, Rodriguez:2023xjd}, or with nontrivial environment \cite{Cardoso:2019upw, Cannizzaro:2024fpz, Barbosa:2026qcv}. Consequently, an observationally measured deviation from vanishing static Love numbers would provide a promising smoking-gun signature of physics beyond the standard paradigm of black holes in four-dimensional GR.

Current observational methods for studying compact objects do not provide direct access to their structure in the vicinity of the expected horizon \cite{Cardoso:2008bp, Konoplya:2017wot}. Instead, many of the observable properties of these objects are primarily sensitive to the region around the photon sphere and the corresponding peak of the effective potential \cite{Abramowicz:2002vt}. Consequently, observations provide only limited information about the structure of the compact object in the near-horizon region. This makes the study of tidal deformability particularly valuable and motivates the investigation of alternative horizonless compact objects, such as gravastars \cite{Mazur:2001fv}, boson stars \cite{Schunck:2003kk, Herdeiro:2021lwl}, ultracompact stars \cite{Abramowicz:1997qk}, wormholes \cite{Morris:1988tu, Lemos:2008cv, Damour:2007ap}, fuzzballs \cite{Mathur:2005zp}, and other proposed candidates \cite{Cardoso:2019rvt}, collectively referred to as black-hole mimickers.

In this work, we focus on wormholes and consider two particular models. The first is a thin-shell wormhole \cite{Visser:1989kg}, constructed by gluing two copies of the Schwarzschild geometry,
\begin{equation}
\text{d}s^2
=-\left(1-\dfrac{r_s}{r}\right)\text{d}t^2
+\dfrac{\text{d}r^2}{1-\frac{r_s}{r}}
+r^2\left(\text{d}\theta^2+\sin^2\theta\text{d}\phi^2\right),
\end{equation}
along the hypersurface $r=r_s(1+\epsilon)$,  $\epsilon\ll 1$, located just outside the Schwarzschild horizon. 
The radial derivatives of the metric are not continuous across the junction surface $r=r_s(1+\epsilon)$.

The second is the Damour-Solodukhin (DS) wormhole \cite{Damour:2007ap},
\begin{equation}
\text{d}s^2
=-\left(1-\dfrac{r_s}{r}+\lambda^2\right)\text{d}t^2
+\dfrac{\text{d}r^2}{1-\frac{r_s}{r}}
+r^2\left(\text{d}\theta^2+\sin^2\theta\text{d}\phi^2\right),
\end{equation}
which can be viewed as a deformation of the Schwarzschild geometry controlled by the small, dimensionless parameter $\lambda$. In the limit $\lambda\rightarrow 0$, the DS wormhole approaches the Schwarzschild black-hole geometry. For any nonzero $\lambda$, the DS geometry is a smooth deformation of the Schwarzschild spacetime, with the metric and all of its derivatives remaining regular and continuous across the would-be horizon.

Within four-dimensional GR, traversable wormholes generally require exotic matter that violates one or more of the standard energy conditions \cite{Morris:1988tu, Visser:1995cc}. In this work, however, we do not address the specific mechanism responsible for supporting or forming the wormholes, nor do we assume a particular modified theory of gravity in which the geometries considered here may arise. In the absence of such a specific gravitational model, we restrict our analysis to scalar perturbations of the backgrounds described above. While scalar perturbations do not capture the full dynamics of gravitational perturbations, we expect them to exhibit qualitatively similar features relevant to the tidal response.

Wormhole solutions can nevertheless arise in alternative theories of gravity \cite{Harko:2013yb, Moraes:2017dbs} and in higher-dimensional GR \cite{Svitek:2016nvm}, where the requirement of exotic stress-energy can be avoided under appropriate conditions. Even within four-dimensional GR, wormhole geometries may be supported by quantum effects, such as Casimir energy \cite{Garattini:2019ivd}, or by the backreaction of quantum conformal fields associated with the conformal anomaly \cite{Berthiere:2017tms}. Two-dimensional models provide further examples \cite{Potaux:2021yan, Potaux:2022uxa, Potaux:2023fwm}, in which quantum backreaction can replace the classical event horizon with a wormhole throat.

Wormholes are particularly interesting because they provide examples of spacetimes with nontrivial topology, in which two asymptotically flat regions are connected by a throat. In the present context, the wormholes considered here can closely mimic the properties of a black hole. As a result, the effective radial potential for scalar perturbations in these backgrounds develops two peaks separated by a finite distance (see Fig.~\ref{fig:effpot}), with the separation scaling as $r_s\ln(1/\lambda)$. This double-barrier structure can lead to distinctive observational signatures, such as a double shadow in asymmetric wormholes \cite{Wielgus:2020uqz, Wang:2020emr, Guerrero:2022qkh, Guerrero:2021pxt, Solodukhin:2025opw}, or, more relevant to the present work, gravitational-wave echoes \cite{Solodukhin:2025opw, Cardoso:2016oxy, Cardoso:2016rao, Hui:2019aox, Abedi:2016hgu}: a sequence of repeated, time-delayed signals following the primary ringdown. The primary signal is governed by black-hole-like QNMs \cite{Cardoso:2016oxy, Hui:2019aox} and can therefore be nearly indistinguishable from the ringdown of a black hole. The QNMs of the wormhole itself \cite{Solodukhin:2025opw, Bueno:2017hyj}, defined by the poles of the retarded Green's function associated with the full effective potential, are instead distinct from the black-hole QNMs. Their spectrum is characterized by long-lived oscillations with small imaginary parts.

The waveform, however, closely resembles that of a black hole. At early times, the signal is dominated by scattering off the first potential barrier, which is essentially identical to the corresponding barrier of a black hole. Only at later times, when the perturbation propagates across the throat, scatters off the second potential barrier, and returns, do deviations from the black-hole waveform become apparent. At this stage, the signal begins to reveal the underlying two-barrier structure of the wormhole. The corresponding time delay is again of order $L_{\text{th}}\sim r_s\ln(1/\lambda)$ in the DS wormhole and $L_{\text{th}}\sim r_s \ln(1/\epsilon)$ in the thin-shell wormhole.

In this work, we extend this picture to the study of dynamical Love numbers. We show that the emergence of a new characteristic length scale, namely the throat length, naturally separates the tidal response into two physically distinct classes: sub-throat modes, whose typical wavelength is much smaller than the throat length, and super-throat modes, whose wavelengths are larger than the throat length, $L_{\rm th}$. The Love numbers associated with these two classes differ in a fundamental way. Super-throat modes characterize the global response of the wormhole and are sensitive to the geometry of both asymptotic regions. This response differs from that which can be inferred by a distant observer over a finite observation time, for whom the sub-throat modes provide the appropriate description. We further show that, similarly to the quasinormal modes of wormholes, the full dynamical Love numbers do not admit a smooth analytic limit to those of the corresponding black holes as the wormhole geometry is continuously deformed toward the black-hole limit. In contrast, the sub-throat Love numbers possess a well-defined black-hole limit and smoothly approach the corresponding Schwarzschild result.

The main novelty of the present paper is the identification of two distinct classes of dynamical Love numbers arising from the interplay between the two-barrier structure and the long throat of the wormhole.

Various black-hole mimickers have been studied in the context of tidal Love numbers \cite{Chakraborty:2026qru}, with most works focusing on their static tidal response \cite{Pani:2015tga, Uchikata:2016qku, Biswas:2026vdp, Giri:2024cks}. These calculations generally yield nonvanishing static Love numbers, suggesting that tidal response may help distinguish black-hole mimickers from classical black holes. A seminal work in this direction is \cite{Cardoso:2017cfl}, where Love numbers were computed for several exotic compact objects, including boson stars, gravastars, and thin-shell wormholes, and a characteristic logarithmic behavior in the black-hole limit was identified. In contrast, the dynamical tidal response of black-hole mimickers has received less attention, although frequency-dependent Love numbers have been investigated for some types of such objects \cite{Chakraborty:2023zed, Nair:2022xfm}.

The paper is organized as follows. In section \ref{sec:DifOfTLNs}, we review the definition of the  Love numbers. In section \ref{sec:tworegims}, we discuss the two classes of modes relevant to our analysis and explain their physical origin. In section \ref{sec:nearfarapp}, we describe the method for computing the Love numbers, based on the separation of the problem into near and far zones.  In sections \ref{sec:SchWH} and \ref{sec:DSWH}, we present the calculation of the Love numbers for both classes of modes in the two wormhole geometries considered in this work. Finally, in section \ref{sec:staticandblack-hole}, we discuss two important limits of the Love numbers: the static limit and the black-hole limit. We summarize in section  \ref{conclusions}.

\section{Definition of tidal Love numbers}\label{sec:DifOfTLNs}
We begin by outlining the basic idea behind the tidal Love numbers. To do so, let us start with their definition in Newtonian gravity. Consider a spherically symmetric body of mass $M$ placed in an external adiabatic tidal field $U_{\text{ext}}=-\sum_{l=2}\mathcal{E}(t)_{i_{1}\dots i_{l}}x^{i_{1}}\dots x^{i_{l}}$, where $\mathcal{E}(t)_{i_1\dots i_l}$ are its multipole moments. In response to this external field, the body becomes deformed and develops induced multipole moments $I(t)_{i_{1}\dots i_{l}}$. Combining all contributions, we obtain the following total Newtonian potential
\begin{equation}
    U^{\text{N}}=-\dfrac{M}{r}-\sum_{l=2}\sum_{m=-l}^{l}Y_{lm}\left[\mathcal{E}_{lm}r^l-\dfrac{I_{lm}}{r^{l+1}}\right],
    \label{NewtonpotentialDecomp1}
\end{equation}
where we have expanded the potential in spherical harmonics. Since the external field is assumed to be weak and adiabatic, linear response theory implies \cite{Chakrabarti:2013xza} that the induced multipole moments $I_{lm}$ are proportional to the external tidal multipoles, possibly with a time delay $\tau$
\begin{equation}
    I_{lm}(t)\approx \lambda_l\mathcal{E}_{lm}(t-\tau).
\end{equation}
The proportionality constants are known as the response coefficients, while the time delay is encoded in the dynamical and dissipative parts of these coefficients. Therefore, the total potential takes the form 
\begin{equation}
    U^{\text{N}}=-\dfrac{M}{r}-\sum_{l=2}^{\infty}\sum_{m=-l}^{l}Y_{lm}\mathcal{E}_{lm}r^l\left[1+k_{lm}\left(\dfrac{R}{r}\right)^{2l+1}\right],
    \label{NewtonpotentialDecomp2}
\end{equation}
where $k_{lm}=-\lambda_{lm}R^{-2l-1}$ are the dimensionless response coefficients, also known as the TLNs, and $R$ is a characteristic size of the object. 
Thus, in Newtonian gravity, the TLNs can be obtained by reading off the coefficient of the $r^{-l-1}$ term in the asymptotic expansion of the Newtonian potential \eqref{NewtonpotentialDecomp2}.

In GR, however, the gravitational potential receives relativistic corrections, and the general potential $U^{\text{GR}}=-\frac{1}{2}(1+g_{00})$ takes the form \cite{Binnington:2009bb, Damour:2009vw, Kol:2011vg, Poisson:2014gka, Yunes:2005ve}
\begin{equation}
    \begin{aligned}
        U^{\text{GR}}=-\dfrac{M}{r}-\sum_{l=2}^{\infty}\sum_{m=-l}^{l}Y_{lm}\mathcal{E}_{lm}r^l
        &\left[\left(1+\sum_{n=1}^{\infty}a_n\left(\dfrac{R}{r}\right)^{n}\right)+\right.
        \\
        &+\left.k_{lm}\left(\dfrac{R}{r}\right)^{2l+1}\left(1+\sum_{n=1}^{\infty}b_n\left(\dfrac{R}{r}\right)^{n}\right)\right].
    \end{aligned}
    \label{GRpotentialDecomp1}
\end{equation}
Here, the first set $\{a_n\}$ describes corrections to the external tidal field, while the second $\{b_n\}$ corresponds to corrections to the induced multipole moments. As is evident from this expression, the definition of TLNs in GR becomes ambiguous because the relativistic correction $a_{2l+1}$ to the external field overlaps with the coefficients $k_{lm}$. Moreover, this definition is generally coordinate-dependent \cite{Gralla:2017djj}.

A rigorous and coordinate-invariant definition of TLNs is provided by the effective field theory (EFT) approach \cite{Charalambous:2021mea, Goldberger:2004jt, Goldberger:2005cd, Rodriguez:2026iot, Ivanov:2022hlo}. In this framework, the compact object is modeled as a point particle moving along its worldline, while all finite-size effects are encoded in higher-derivative operators. The coefficients multiplying these operators, known as Wilson coefficients, are directly related to the TLNs.
 Their values are determined by matching the EFT to the corresponding microscopic calculation. Thus, EFT provides an unambiguous and coordinate-independent definition of TLNs.

Another widely used technique for defining TLNs is to treat the angular momentum $l$ as an arbitrary real parameter rather than restricting it to positive integer values \cite{Kol:2011vg, LeTiec:2020bos}. This analytic continuation avoids the ambiguity in separating the $r^{-l-1}$ and $r^{l}$ branches of the solution \eqref{GRpotentialDecomp1}, thereby allowing the TLNs to be read off unambiguously.\footnote{This procedure may also be motivated by EFT. In dimensional regularization, the angular momentum $l$ naturally appears in the combination $\frac{l}{D-3}$, where $D$ is the number of spacetime dimensions.}

In this work, we adopt this technique to define the TLNs of the wormholes under consideration. 
 Since no underlying gravitational theory in which these wormholes arise as exact solutions is currently known, and for simplicity, we restrict our analysis to scalar perturbations. The ideas discussed above carry over directly to the scalar case. Although simpler, this setup already captures the main properties of the Love numbers and is expected to provide useful insight into the corresponding gravitational TLNs.\footnote{Strictly speaking, the term Love numbers is often reserved for gravitational perturbations, while the corresponding quantities for a scalar field are referred to as scalar tidal numbers. Throughout this paper, however, we will also refer to the scalar quantities as TLNs.} 

\section{Modes of interest}\label{sec:tworegims}

\begin{figure}[htbp]
    \centering
    \includegraphics[width=0.7\textwidth]{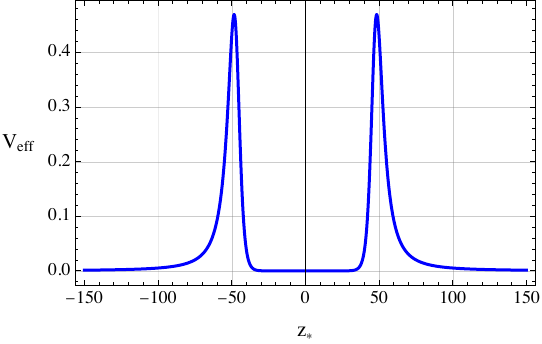}
    \caption{Effective potential \eqref{DSWHeffectivepotential} of the DS wormhole in the tortoise coordinate \eqref{DSWHtortoisecoordinate} for $l=3$.}
    \label{fig:effpot}
\end{figure}

The wormholes considered in this work consist of two asymptotically flat universes connected by a long tube, which we call the throat. Another characteristic feature of these wormholes is the presence of two peaks in the effective potential, one in each universe, for both geodesic motion and field perturbations. Fig.~\ref{fig:effpot} 
shows the effective potential for scalar perturbations in the DS wormhole (see section \ref{sec:DSWH}). As a result, the problem involves two characteristic length scales: the throat radius $r_\text{th}$ (analogous to the horizon radius of a black hole) and the throat length $L_{\text{th}}$ (which can be defined as the distance between the two peaks of the effective radial potential). The throat length depends logarithmically on the parameter $\mu$, which measures the deviation of the wormhole geometry from that of a black hole
\begin{equation}
    L_{\text{th}}=r_\text{th}\ln\dfrac{1}{\mu}.
    \label{throatlength}
\end{equation}
Since this parameter is expected to arise from quantum effects, one naturally expects $\mu\ll 1$. Consequently, $r_\text{th}\ll L_{\text{th}}$, that is, the throat is much longer than its radius. When considering dynamical (wave) processes in this spacetime, it is useful to distinguish three distinct classes of  modes.

It is worth emphasizing that our Love numbers are computed in the frequency domain, assuming a monochromatic external tidal perturbation. Therefore, although they characterize the dynamical response of the system, such calculations do not directly describe its evolution in time. A physical perturbation, however, has a finite duration and can be represented as a superposition of monochromatic waves. Its time-domain evolution therefore introduces characteristic timescales that can be compared with the propagation times associated with the wormhole geometry, in particular the time required for a wave to reach and probe the throat. Accordingly, when discussing different frequency ranges below, we associate a mode of characteristic frequency $\omega$ with a characteristic timescale of order $\frac{1}{\omega}$.  This provides a useful time-domain interpretation of the frequency-domain results in terms of whether the perturbation has sufficient time to reach the throat and, on longer timescales, to probe its global structure. Throughout the paper, we will use this correspondence when moving between the frequency-domain calculation and its time-domain interpretation.

For \textbf{short-wavelength modes},  the wavelength $\lambdabar$ is much smaller than all characteristic length scales $\lambdabar=\frac{2\pi}{\omega}\ll r_\text{th}\ll L_{\text{th}}$. Such waves are sensitive to the fine structure of the geometry and can therefore probe small-scale features. 

In the second case, $r_\text{th}\ll\lambdabar= \frac{2\pi}{\omega}\ll L_{\text{th}}$, the wavelength is much larger than the throat radius but still much smaller than the throat length, or equivalently $\omega r_\text{th}\ll 1 \ll \omega L_{\text{th}}$. Although this is already a low-frequency regime, we refer to these as the \textbf{sub-throat modes} to emphasize their distinctive physical properties. Such waves no longer resolve the small-scale structure of the geometry, but they can still detect the presence of a long throat connecting the two universes. 

This case can also be understood from a time-domain perspective. Consider an observer in one asymptotically flat universe who sends a wave toward the wormhole and monitors the scattered signal. Suppose the observation time is long enough for the wave to interact with the effective potential barrier in the observer's own universe, but not long enough for the signal to travel through the throat, scatter from the potential barrier in the opposite universe, and return. The characteristic time required for this round trip is known as the echo time $\Delta t_{\text{echo}}$, and it is directly related to the throat length $\Delta t_{\text{echo}}=2L_{\text{th}}$. In this situation, the observer may be unable to distinguish the wormhole from a black hole. This is closely related to the well-known behavior of wormhole ringdown signals \cite{Cardoso:2016oxy, Cardoso:2016rao, Hui:2019aox, Abedi:2016hgu}, where the initial waveform closely resembles that of a black hole, while the differences appear only in the subsequent echo signals. 

In the third regime $r_{\text{th}}\ll L_{\text{th}}\ll \lambdabar=\frac{2\pi}{\omega}$, the wavelength is much larger than all characteristic length scales, or equivalently $\omega r_{\text{th}}\ll \omega L_{\text{th}}\ll 1$. This is also a low-frequency regime. However, waves are now too long to resolve even the presence of the throat itself. Instead, they are sensitive only to the asymptotic structure of the spacetime, namely the centrifugal potential barriers in the asymptotically flat universes. For this reason, we refer to these modes as the \textbf{super-throat modes}.\footnote{We emphasize that the terms "sub-throat" and "super-throat" refer to the wavelength relative to the throat length, not to the spatial localization of the modes.}

From the time-domain point of view, this regime corresponds to observations performed over sufficiently long times that the observer receives signals returning from the opposite universe, namely the echo signals. 

Our main goal is to study the dynamical TLNs of wormholes under an external adiabatic perturbation. Therefore, we will focus only on the sub-throat and super-throat modes. For the high-frequency modes, the response becomes intrinsically nonlocal.\footnote{This is also related to the method used to compute the dynamical TLNs. As discussed in the next section, our calculation relies on the matched near/far zones approximation, which is valid only in the low-frequency limit $\omega r_{\text{th}}\ll 1$.}

Both low-frequency regimes are physically important. The super-throat modes describe the late-time dynamics, when the observer has already received the echo signals and the quasinormal modes (QNMs) of the wormhole itself become relevant \cite{Solodukhin:2025opw}. In contrast, the sub-throat modes allow us to understand the limit $\mu\rightarrow 0$, in which the wormhole's throat becomes increasingly long. As a result, fewer and fewer waves are able to reach the opposite universe within a finite observation time. Consequently, the super-throat modes gradually become irrelevant, and the TLNs are entirely determined by the sub-throat modes. As we will show, this behavior correctly reproduces the TLNs of black holes in the limit $\mu\rightarrow 0$.

As discussed above, for sub-throat modes, the wormhole appears as a black hole. An observer located far from the wormhole, but who does not wait long enough to detect the arrival of the echo signal, observes only the waves scattered by the peak of the effective potential in their own universe. Consequently, the transmission coefficient $\Gamma(\omega)$, defined as
\begin{equation}
    \Box\varphi(x)=0\quad\Rightarrow\quad\left.\varphi(x)\right|_{|\mathbf{x}|\rightarrow\infty}\sim A_{\text{in}}\dfrac{e^{-i\omega|\mathbf{x}|}}{|\mathbf{x}|}+A_{\text{out}}\dfrac{e^{i\omega|\mathbf{x}|}}{|\mathbf{x}|}\quad\Rightarrow\quad\Gamma(\omega)=1-\dfrac{|A_{\text{out}}|^2}{|A_{\text{in}}|^2},
\end{equation}
is determined solely by the centrifugal part of the potential, which falls off as $\sim r^{-2}$ \cite{Unruh:1976fm}. For each multipole $l$, it contributes a factor $\omega^{2l}$, while the geometric size of the object provides an additional factor of $\omega^2$.\footnote{For $l=0$, the effective potential contains no centrifugal barrier. In this case, the absorption cross section $\sigma_0$ is determined only by the effective area of the absorbing surface, and therefore $\Gamma_0(\omega)\sim\omega^2$, since $\Gamma_l(\omega)\sim \omega^2\sigma_{l}(\omega)$.} For a black hole, the low-frequency behavior is \cite{Das:1996we, Starobinsky:1973aij, Starobinsky:1973aij2}
\begin{equation}
    \Gamma_l(\omega)\sim (\omega r_s)^{2l+2}.
    \label{subthroatprediction}
\end{equation}

Since sub-throat modes interact with only one of the two potential barriers , we expect the same low-frequency scaling for $\Gamma_l(\omega)$ \cite{Freivogel:2026ujn}. Furthermore, because the leading-order imaginary part of the Love numbers is related to $\Gamma_l(\omega)$ through
\begin{equation}
    \Im k_{\text{Love}}=\dfrac{(2l)!(2l+1)!}{2(l!)^2}\dfrac{\Gamma_l(\omega)}{(2\omega r_s)^{2l+1}},
    \label{imloveandtranscoef}
\end{equation}
we expect the same low-frequency behavior $\Im k_{\text{Love}}\sim \omega r_s$ as for the dynamical Love numbers of a black hole \cite{Combaluzier--Szteinsznaider:2025eoc, Kobayashi:2025vgl}.

The situation is qualitatively different for super-throat modes. Such modes probe the entire global geometry of the wormhole, and the observer waits long enough to receive waves reflected from the opposite universe. Consequently, the observed signal is scattered by both centrifugal potential barriers, one in each universe. Since each barrier contributes a factor of $\omega^{2l}$ at low frequencies, the transmission coefficient is expected to scale as
\begin{equation}
    \Gamma_l(\omega)\sim \omega^{4l+2}.
\end{equation}
Eq.~\eqref{imloveandtranscoef} then implies that, for super-throat modes, the imaginary part of the Love numbers should be 
\begin{equation}
    \Im k_{\text{Love}} \sim (\omega r_s)^{2l+1}.
    \label{superthroatimlove}
\end{equation}
In particular, this explains why the imaginary part of the Love numbers for the wormholes considered here exhibits a higher power of the frequency $\omega$ in the calculations below. This behavior is markedly different from that of the dynamical Love numbers of black holes.

\section{Near/Far zones approximation}\label{sec:nearfarapp}
In this work, we consider a spherically symmetric wormhole metric  which connects two asymptotically flat regions.
 The metric in each region can be written in the following form
\begin{equation}
    \text{d}s^2=-\dfrac{\bar{\Delta}(r)}{r^2}\text{d}t^2+\dfrac{r^2}{\Delta(r)}\text{d}r^2+r^2\left(\text{d}\theta^2+\sin^2\theta \text{d}\phi^2\right).
    \label{commonmetric}
\end{equation} 
To satisfy the condition of asymptotic flatness, the metric functions must behave at spatial infinity as $\Delta(r)\sim r^2$ and $\bar{\Delta}(r)\sim r^2$ (the coefficients in front of $r^2$ can be set to unity by an appropriate redefinition of the coordinates $t$ and $r$). 

Throughout this paper, we study a minimally coupled scalar field propagating on the background geometry \eqref{commonmetric}
\begin{equation}
    \nabla_{\mu}\nabla^{\mu}\varphi(x)=0\quad\Rightarrow\quad
    \sqrt{\dfrac{\Delta}{\bar{\Delta}}}\partial_{r}\left(\sqrt{\Delta\bar{\Delta}}\partial_{r}R(r)\right)+\left(\dfrac{r^4\omega^2}{\bar{\Delta}}-l(l+1)\right)R(r)=0,
    \label{scalreq}
\end{equation}
where we have decomposed the scalar field as $\varphi(x)=\sum_{lm}e^{-i\omega t}R(r)Y_{lm}(\theta,\phi)$. Depending on the specific form of the metric functions, this equation may possess different sets of singular points. However, the singular point at infinity is universal and is always an irregular singularity as a consequence of asymptotic flatness. The presence of an irregular singular point significantly complicates the analysis of the equation. For the Schwarzschild and Kerr geometries, several methods have been developed to solve equations of this type. These include the Mano-Suzuki-Takasugi (MST) expansion \cite{Mano:1996vt, Mano:1996mf}, which expresses the solution as a series of hypergeometric functions, exact solutions in terms of the confluent Heun functions \cite{Fiziev:2005ki, Bonelli:2021uvf} and recently derived connection formulas for these functions \cite{Bonelli:2022ten, Lisovyy:2022flm}, and the matched near/far zone approximation \cite{Unruh:1976fm, Starobinsky:1973aij, Starobinsky:1973aij2, Hui:2022vbh}.

In this work, we employ the latter approach for the wormhole geometry. This approximation requires the low-frequency limit $\omega r_0\ll 1$ and consists of dividing the spacetime into two regions. The \textit{near zone} is defined by 
\begin{equation}
    r_0\leq r\ll\omega^{-1},
    \label{nearzone}
\end{equation}
where $r_0$ is the characteristic length scale of the object. In our case, $r_0=r_{\text{th}}$ corresponds to the size of the wormhole throat. The \textit{far zone} is defined by 
\begin{equation}
    r\gg r_0,
    \label{farzone}
\end{equation}
where the spacetime is almost asymptotically flat. These two regions overlap in the \textit{intermediate zone} 
\begin{equation}
    r_0\ll r\ll \omega^{-1}.
    \label{intermediatezone}
\end{equation}
Within this overlap, the near and far zone solutions can be matched. Since we consider a wormhole geometry connecting two asymptotically flat regions, the spacetime naturally consists of one near zone, covering the throat and its vicinity, and two far zones, one in each asymptotic infinity. Consequently, there are two intermediate zones, one in each asymptotic region. It is important to note that the definition of the TLNs requires the scalar field solution to be expressed as a linear combination of the growing mode $r^l$ and the decaying mode $r^{-l-1}$, similar to \eqref{GRpotentialDecomp1}. Therefore, the TLNs are defined with respect to the intermediate region, where such an expansion is valid. 

In the \textit{near zone}, our goal is to isolate the dominant singular contribution to the coefficients of the differential equation. Mathematically, this amounts to replacing the original equation by an auxiliary one in which the point at infinity becomes a regular singular point, making the equation considerably easier to analyze. To achieve this, we first introduce the dimensionless coordinate $z=\frac{r-r_0}{r_0}$, where the value $r_0$ depends on the choice of the origin of the radial coordinate. In terms of $z$, the radial equation takes the form
\begin{equation}
    \sqrt{\dfrac{\Delta(r(z))}{\bar{\Delta}(r(z))}}\dfrac{1}{r_0^2}\partial_{z}\left(\sqrt{\Delta(r(z))\bar{\Delta}(r(z))}\partial_{z}R(z)\right)+\left(\dfrac{r_0^4(1+z)^4\omega^2}{\bar{\Delta}(z)}-l(l+1)\right)R(z)=0.
\end{equation}
The point at infinity remains an irregular singularity because of the terms proportional to $\omega$. To identify which contribution is responsible for this irregular behavior, we introduce a formal parameter $\alpha$ \cite{Charalambous:2021mea} as follows: $r_0^4\omega^2(1+\alpha z)^4$. Although its physical value is $\alpha=1$, we temporarily keep it arbitrary. After making a second change of variable $z\rightarrow x=\frac{1}{z}$, the asymptotic form of the equation as $x\rightarrow 0$ is
\begin{equation}
    \dfrac{\text{d}^2}{\text{d}x^2}R(x)+\left((\omega r_0)^2\left(1+\dfrac{\alpha}{x}\right)^4-\dfrac{l(l+1)}{x^2}\right)R(x)=0.
\end{equation}
Here we have used the asymptotic forms of $\Delta(r)\sim r^2\Rightarrow \Delta(x)\sim \frac{r_0^2}{x^2}$ and $\bar{\Delta}(r)\sim r^2\Rightarrow \bar{\Delta}(x)\sim \frac{r_0^2}{x^2}$. From this expression, it is clear that the parameter $\alpha$ is responsible for making the point at infinity an irregular singularity. Therefore, to obtain the near zone equation, we set $\alpha=0$. This procedure is equivalent to replacing the term $\omega^2r^4$ in the original equation \eqref{scalreq} by $\omega^2r_0^4$, where $r_0$ is a convenient characteristic length scale. 

As a result, the radial equation in the near zone takes the following form 
\begin{equation}
    \sqrt{\dfrac{\Delta}{\bar{\Delta}}}\partial_{r}\left(\sqrt{\Delta\bar{\Delta}}\partial_{r}R(r)\right)+\left(\dfrac{r_0^4\omega^2}{\bar{\Delta}}-l(l+1)\right)R(r)=0.
    \label{nearzonescalarequation}
\end{equation}
In contrast to the original equation, the point at infinity is now a regular singular point.

In the \textit{far zone}, $r\gg r_0$. Therefore, the scalar-field equation \eqref{scalreq} reduces to the spherical Bessel equation
\begin{equation}
    \partial_r\left(r^2\partial_rR(r)\right)+\left(\omega^2r^2-l(l+1)\right)R(r)=0,
\end{equation}
as a consequence of asymptotic flatness. Therefore, the general solution has the following form in both universes
\begin{equation}
    R^{\text{(r,l)}}(r)=\sqrt{\dfrac{\pi}{2\omega r}}\left[B^{\text{(r,l)}}_1J_{l+\frac{1}{2}}(\omega r)+B^{\text{(r,l)}}_2(-1)^{l+1}J_{-l-\frac{1}{2}}(\omega r)\right].
    \label{solfarzone}
\end{equation}
Here, $R^{\text{(r)}}(r)$ and $R^{\text{(l)}}(r)$ are solutions in the right and left universes, respectively. Using the known asymptotics of the Bessel functions, we write the solution in the intermediate zone ($\omega r\ll 1$), where it will later be matched to the near-zone solution
\begin{equation}
    R^{\text{(r,l)}}(r)\sim B^{\text{(r,l)}}_1\dfrac{4^l\Gamma(l+1)}{\Gamma(2l+2)}\left(\dfrac{\omega r_s}{2}\right)^l\left(\dfrac{r}{r_s}\right)^l-B^{\text{(r,l)}}_2\dfrac{\Gamma(2l+1)}{2\Gamma(l+1)4^l}\left(\dfrac{\omega r_s}{2}\right)^{-l-1}\left(\dfrac{r}{r_s}\right)^{-l-1}.
    \label{farzoneintermatch}
\end{equation}

We also write, for the left universe only, the expansion at infinity ($\omega r\gg 1$), since the boundary condition will be imposed there
\begin{equation}
    R^{\text{(l)}}(r)\sim\dfrac{-B_2^{\text{(l)}}-iB_1^{\text{(l)}}}{2}e^{-i\frac{l\pi}{2}}\dfrac{e^{i\omega r}}{\omega r}+\dfrac{-B_2^{\text{(l)}}+iB_1^{\text{(l)}}}{2}e^{i\frac{l\pi}{2}}\dfrac{e^{-i\omega r}}{\omega r}
\end{equation}
In all cases considered below, the tortoise coordinate in the left universe has the following asymptotic relation to the usual radial coordinate: $z_{\ast}\sim -r$. Therefore, the solution can be rewritten in the following form
\begin{equation}
    R^{\text{(l)}}(r)\sim\dfrac{-B_2^{\text{(l)}}-iB_1^{\text{(l)}}}{2}e^{-i\frac{l\pi}{2}}\dfrac{e^{-i\omega z_{\ast}}}{\omega r}+\dfrac{-B_2^{\text{(l)}}+iB_1^{\text{(l)}}}{2}e^{i\frac{l\pi}{2}}\dfrac{e^{i\omega z_{\ast}}}{\omega r}
\end{equation}
Since we assume that there is no wave reflected from spatial infinity in the left universe, the coefficients must satisfy the following relation
\begin{equation}
    B^{\text{(l)}}_2=iB^{\text{(l)}}_1.
    \label{conditioninleftuniversedyn}
\end{equation}
This will be the main condition used below to relate   the two basis solutions in the near zone.

It is important to note that all subsequent calculations performed within the near/far zone approximation are valid only to leading order in $\omega r_0$.

\section{Thin-shell Schwarzschild wormhole}\label{sec:SchWH}
\subsection{Wormhole geometry}\label{sec:schwormholegeometry}

One of the simplest wormhole models can be constructed by gluing two identical copies of the Schwarzschild metric \cite{Visser:1989kg}
\begin{equation}
    \text{d}s^2=-\left(1-\dfrac{r_s}{r}\right)\text{d}t^2+\dfrac{\text{d}r^2}{1-\frac{r_s}{r}}+r^2\left(\text{d}\theta^2+\sin^2\theta\text{d}\phi^2\right)
\end{equation}
across hypersurfaces defined by $r=r_0>r_s$. The resulting spacetime is geodesically complete and contains neither an event horizon nor a spacetime singularity. The throat of the wormhole is located at $r=r_0$. 

As a consequence of the junction conditions \cite{Israel:1966rt}, the stress-energy tensor vanishes everywhere except at the throat. On the hypersurface corresponding to the throat, there is a thin shell of matter with surface energy density and surface pressure given by
\begin{equation}
    \sigma = -\dfrac{1}{2\pi r_0}\sqrt{1-\dfrac{r_s}{r_0}},\quad p=\dfrac{1}{4\pi r_0}\dfrac{1-\frac{r_s}{2r_0}}{\sqrt{1-\frac{r_s}{r_0}}}.
\end{equation}
These expressions imply that the weak and dominant energy conditions are violated, while the null and strong energy conditions are satisfied for $r_0<\frac{3}{2}r_s$. 

In what follows, we assume that the matching surface is located at 
\begin{equation}
    r_0=r_s(1+\epsilon),\quad\epsilon\ll 1.
\end{equation}
The parameter $\epsilon$ characterizes the deformation from a black hole geometry to a wormhole geometry. In this case, the throat length \eqref{throatlength} is given by $L_{\text{th}}=r_s\ln\frac{1}{\epsilon}$. We also introduce the tortoise coordinate, defined by 
\begin{equation}
    \dfrac{\text{d}z_{\ast}}{\text{d}r}=\pm\dfrac{r}{r-r_s},
    \label{SchWHtortoisecoordinate}
\end{equation}
where the $+$ sign corresponds to the right universe and the $-$ sign to the left universe. Without loss of generality, we choose $z_{\ast}(r_0)=0$. Near the throat, the tortoise coordinate behaves as 
\begin{equation}
    z_{\ast}\approx \pm r_s\ln\dfrac{r-r_s}{r_s\epsilon}\quad\Rightarrow\quad r-r_s\approx r_s\epsilon e^{\pm z_{\ast}/r_s}.
    \label{SchWHtortnearzone}
\end{equation}

\subsection{Scalar field equation}\label{sec:SchWHscalarfieldequation}

For the Schwarzschild metric $\Delta(r)=\bar{\Delta}(r)=r(r-r_s)$, and therefore the equation for the radial part of the scalar field takes the following form
\begin{equation}
    \dfrac{\text{d}^2}{\text{d}r^2}R(r)+\left(\dfrac{1}{r}+\dfrac{1}{r-r_s}\right)\dfrac{\text{d}}{\text{d}r}R(r)+\dfrac{\omega^2r^4-l(l+1)r(r-r_s)}{r^2(r-r_s)^2}R(r)=0.
    \label{SchWHscalareqn}
\end{equation}
Introducing the function $\Psi(z_{\ast})=r(z_{\ast})R(z_{\ast})$ and rewriting the equation in terms of the tortoise coordinate \eqref{SchWHtortoisecoordinate}, we obtain a Schrödinger-like equation with an effective potential
\begin{equation}
    \dfrac{\text{d}^2}{\text{d}z_{\ast}^2}\Psi+\left(\omega^2-V_l\right)\Psi=0,\quad V_l(r)=\left(1-\dfrac{r_s}{r}\right)\left(\dfrac{r_s}{r^3}+\dfrac{l(l+1)}{r^2}\right).
    \label{SchWHeffpot}
\end{equation}
Note that at the throat $r_0=r_s(1+\epsilon)$ the effective potential does not vanish and takes the following value $V_l(r_0)=\frac{\epsilon}{r_s^2(1+\epsilon)^3}\left(\frac{1}{1+\epsilon}+l(l+1)\right)=\frac{\epsilon}{r_s^2}\left(1+l(l+1)\right)+\mathcal{O}(\epsilon^2)$. Consequently, waves can propagate inside the throat only if $\omega^2>V_l(r_0)$. The radial equation \eqref{SchWHscalareqn} has three singular points: two regular singular points at $r=0$ and $r=r_s$, and one irregular singular point at $r=\infty$. This equation can be reduced to the confluent Heun equation \cite{Fiziev:2005ki, Bonelli:2021uvf}.

As described in section \ref{sec:nearfarapp}, we apply the replacement $\omega^2r^4\rightarrow \omega^2r_s^4$ in the last term of Eq.~\eqref{SchWHscalareqn}. This yields the radial equation in the near zone approximation
\begin{equation}
    \dfrac{\text{d}^2}{\text{d}r^2}R(r)+\left(\dfrac{1}{r}+\dfrac{1}{r-r_s}\right)\dfrac{\text{d}}{\text{d}r}R(r)+\dfrac{\omega^2r_s^4-l(l+1)r(r-r_s)}{r^2(r-r_s)^2}R(r)=0.
    \label{SchWHscalareqnnearzone}
\end{equation}
In this form, all three singular points $r=0$, $r=r_s$ and $r=\infty$ are regular singular points. Therefore, the solution in both universes can be expressed in terms of hypergeometric functions
\begin{equation}
    \begin{aligned}
        R^{\text{(R,L)}}(r)=\left(\dfrac{r}{r_s}\right)^l
        &\left[A_1^{\text{(R,L)}}\left(1-\dfrac{r_s}{r}\right)^{-i\omega r_s}{}_2F_1\left(-l,-l-2i\omega r_s,1-2i\omega r_s;1-\dfrac{r_s}{r}\right)+\right.
        \\
        &\left.+A_2^{\text{(R,L)}}\left(1-\dfrac{r_s}{r}\right)^{i\omega r_s}{}_2F_1\left(-l,-l+2i\omega r_s,1+2i\omega r_s;1-\dfrac{r_s}{r}\right)\right].
\end{aligned}
\label{SchWHnearzonesolution}
\end{equation}

For a black hole, the two solutions proportional to the coefficients $A_1$ and $A_2$ represent ingoing and outgoing waves at the horizon $r=r_s$, respectively. Since the horizon is a perfectly absorbing surface, one imposes the boundary condition $A_2=0$. For the wormhole, however, neither of these solutions corresponds to a purely ingoing or purely outgoing wave at the throat (see appendix \ref{app:SCHwavebasis} for details). We therefore keep both solutions at this stage.

Next, we use the standard connection formulas for the hypergeometric function to relate the solutions near $r=r_s$ to those at $r=\infty$. This gives the following asymptotic solution in the intermediate zones in both universes
\begin{equation}
    \begin{aligned}
        R^{\text{(R,L)}}(r)
        &=\left(1-\dfrac{r_s}{r}\right)^{-i\omega r_s}\Bigg[\left(A_1^{\text{(R,L)}}\dfrac{\Gamma(1-2i\omega r_s)}{\Gamma(l+1-2i\omega r_s)}+A_2^{\text{(R,L)}}\dfrac{\Gamma(1+2i\omega r_s)}{\Gamma(l+1+2i\omega r_s)}\right)\dfrac{\Gamma(2l+1)}{\Gamma(l+1)}\left(\dfrac{r}{r_s}\right)^lZ^{\text{source}}(r)+
        \\
        &+\left(A_1^{\text{(R,L)}}\dfrac{\Gamma(1-2i\omega r_s)}{\Gamma(-l-2i\omega r_s)}+A_2^{\text{(R,L)}}\dfrac{\Gamma(1+2i\omega r_s)}{\Gamma(-l+2i\omega r_s)}\right)\dfrac{\Gamma(-2l-1)}{\Gamma(-l)}\left(\dfrac{r}{r_s}\right)^{-l-1}Z^{\text{response}}(r)\Bigg],
    \end{aligned}
    \label{SchWHasymptoticsolution}
\end{equation}
where we have introduced the following source and response functions
\begin{equation}
    \begin{aligned}
        &Z^{\text{source}}(r)=\left(1-\dfrac{r_s}{r}\right)^l{}_2F_1\left(-l,-l+2i\omega r_s,-2l;\dfrac{r_s}{r_s-r}\right)
        \\
        &Z^{\text{response}}(r)=\left(1-\dfrac{r_s}{r}\right)^{-l-1}{}_2F_1\left(l+1,l+1+2i\omega r_s,2l+2;\dfrac{r_s}{r_s-r}\right).
    \end{aligned}
    \label{SchWHsourceresposesplit}
\end{equation}
Note that both functions approach unity as $r\rightarrow\infty$, and they are mapped into one another under $l\leftrightarrow -l-1$. Throughout this analysis, we treat $l$ as a non-integer in order to avoid the potential divergences that arise for positive integer values of $l$. 

\subsection{The super-throat modes}\label{sec:SchWHsuperthroatmodes}

We begin by computing the Love numbers for the super-throat modes. Since both universes contribute to the response in this case, the solutions in the right and left universes \eqref{SchWHnearzonesolution} must be matched at the throat $r_0=r_s(1+\epsilon)$. Because the effective potential \eqref{SchWHeffpot} is continuous at the throat and contains no delta-function terms, we impose the following matching conditions
\begin{equation}
    \begin{cases}
        \Psi^{\text{(R)}}(z_{\ast}=0) = \Psi^{\text{(L)}}(z_{\ast}=0)
        \vspace{0.2cm}
        \\
        \dfrac{\text{d}}{\text{d}z_{\ast}}\Psi^{\text{(R)}}(z_{\ast}=0) = \dfrac{\text{d}}{\text{d}z_{\ast}}\Psi^{\text{(L)}}(z_{\ast}=0)
    \end{cases},
\end{equation}
where $\Psi^{\text{(R)}}$ and $\Psi^{\text{(L)}}$ denote the solution in the right and left universes, respectively. These matching conditions are equivalent to requiring that the scalar field flux $\mathcal{F}=\frac{1}{2i}\left(\Psi^{\dagger}\partial_{z_{\ast}}\Psi-\Psi\partial_{z_{\ast}}\Psi^{\dagger}\right)$ is conserved across the throat: $\mathcal{F}^{(\text{R})}=\mathcal{F}^{(\text{L})}$.

Using these matching conditions, one obtains rather cumbersome relations between the coefficients $\left(A_1^{\text{(R)}},A_2^{\text{(R)}}\right)$ and $\left(A_1^{\text{(L)}},A_2^{\text{(L)}}\right)$ in the two universes. We do not display their explicit form here. Since our aim is to compute the Love numbers only to first order in $\epsilon$, we instead expand these relations and retain terms up to this order
\begin{equation}
    \begin{aligned}
        A_1^{\text{(R)}}=\epsilon^{2i\omega r_s}A_2^{\text{(L)}}\left(1-i\epsilon\dfrac{l(l+1)+1+2\omega^2r_s^2}{\omega r_s}\right)-iA_1^{\text{(L)}}\epsilon\dfrac{l(l+1)+1+4\omega^2r_s^2}{\omega r_s+4\omega^3r_s^3}
        \\
        A_2^{\text{(R)}}=\epsilon^{-2i\omega r_s}A_1^{\text{(L)}}\left(1+i\epsilon\dfrac{l(l+1)+1+2\omega^2r_s^2}{\omega r_s}\right)+iA_2^{\text{(L)}}\epsilon\dfrac{l(l+1)+1+4\omega^2r_s^2}{\omega r_s+4\omega^3r_s^3}.
    \end{aligned}
    \label{SchWHrelationbetweencoeff}
\end{equation}
As can be seen, the resulting expressions also contain factors of the form $\epsilon^{\pm 2i\omega r_s}$. These factors arise from the oscillatory behavior of the solutions near the throat \eqref{SchWHnearzonesolution}, together with the relation between the radial coordinate $r$ and the tortoise coordinate \eqref{SchWHtortnearzone}.

We must now impose the condition that there is no reflected wave at infinity in the left universe, since the source of the external tidal field is assumed to be localized in the right universe. To apply this condition, we first match the near zone solution in the left universe \eqref{SchWHasymptoticsolution} to the corresponding far zone solution \eqref{farzoneintermatch} in the intermediate zone. We then impose the no-reflection condition at spatial infinity in the left universe \eqref{conditioninleftuniversedyn}. This gives the following relation between the coefficients in the left universe
\begin{equation}
    \dfrac{A_1^{\text{(L)}}}{A_2^{\text{(L)}}}=-\dfrac{\Gamma(1+2i\omega r_s)\Gamma(l+1-2i\omega r_s)}{\Gamma(1-2i\omega r_s)\Gamma(l+1+2i\omega r_s)}\underbrace{\dfrac{1-i2^{2l}(\omega r_s)^{2l+1}\frac{\Gamma^3(l+1)}{\Gamma(2l+2)\Gamma^2(2l+1)}\frac{\Gamma(-2l-1)\Gamma(l+1+2i\omega r_s)}{\Gamma(-l)\Gamma(-l+2i\omega r_s)}}{1-i2^{2l}(\omega r_s)^{2l+1}\frac{\Gamma^3(l+1)}{\Gamma(2l+2)\Gamma^2(2l+1)}\frac{\Gamma(-2l-1)\Gamma(l+1-2i\omega r_s)}{\Gamma(-l)\Gamma(-l-2i\omega r_s)}}}_{\kappa}
    \label{SchWHkappa}
\end{equation}

The Love numbers are defined in the intermediate zone of the right universe, where the solution takes the form 
\begin{equation}
    \begin{aligned}
        R^{\text{(R)}}(r)
        &\sim\left(1-\dfrac{r_s}{r}\right)^{-i\omega r_s}\left(\dfrac{r}{r_s}\right)^l\Bigg[Z^{\text{source}}(r)+k_{\text{Love}}\left(\dfrac{r_s}{r}\right)^{2l+1}Z^{\text{response}}(r)\Bigg],
    \end{aligned}
\end{equation}
with 
\begin{equation}
    k_{\text{Love}}=\dfrac{\Gamma(-2l-1)\Gamma(l+1)\Gamma(l+1-2i\omega r_s)}{\Gamma(2l+1)\Gamma(-l)\Gamma(-l-2i\omega r_s)}\dfrac{1-\frac{A^{\text{(R)}}_2}{A^{\text{(R)}}_1}\frac{\Gamma(1+2i\omega r_s)\Gamma(l+1-2i\omega r_s)}{\Gamma(1-2i\omega r_s)\Gamma(l+1+2i\omega r_s)}}{1+\frac{A^{\text{(R)}}_2}{A^{\text{(R)}}_1}\frac{\Gamma(1+2i\omega r_s)\Gamma(l+1-2i\omega r_s)}{\Gamma(1-2i\omega r_s)\Gamma(l+1+2i\omega r_s)}}
    \label{SchWHLove}
\end{equation}
denoting the Love numbers.

Let us clarify the applicability of the matched near/far zone approximation. Although this method is used together with the low-frequency approximation $\omega r_s\ll 1$, naively discarding all terms beyond linear order in $\omega r_s$ in the final expression \eqref{SchWHLove} would be incorrect, as it would remove the leading nontrivial imaginary contribution. The correct procedure is to retain the leading contribution in $\omega r_s$ arising separately from both the near and far regions. In particular, one cannot simply set $\kappa=1$ \eqref{SchWHkappa} on the grounds that it contains higher powers of $\omega r_s$. The combination $(\omega r_s)^{2l+1}$, which appears in both the numerator and denominator of $\kappa$, represents the leading order contribution from the far zone \eqref{farzoneintermatch} and therefore cannot be neglected. Accordingly, $\kappa$ must be expanded while retaining its leading nontrivial dependence on $\omega r_s$
\begin{equation}
    \kappa=1-i2^{2l}(\omega r_s)^{2l+1}\frac{\Gamma(-2l-1)\Gamma^3(l+1)}{\Gamma(-l)\Gamma(2l+2)\Gamma^2(2l+1)}\left(\frac{\Gamma(l+1+2i\omega r_s)}{\Gamma(-l+2i\omega r_s)}-\frac{\Gamma(l+1-2i\omega r_s)}{\Gamma(-l-2i\omega r_s)}\right)+\mathcal{O}\left((\omega r_s)^{2l+3}\right).
    \label{SchWHkappaexp}
\end{equation}
Therefore, the order of the expansion used above is sufficient for the present calculation. 

We now take the limit in which $l$ approaches positive integer values by making the shift $l\rightarrow l-\delta$. As in the Schwarzschild black hole case, no divergence appears in \eqref{SchWHLove}: the poles of the two gamma functions $\Gamma(-2l-1)$ and $\Gamma(-l)$ cancel each other.\footnote{Although the characteristic exponents of the two basis solutions of Eq.~\eqref{SchWHscalareqnnearzone} at $r=\infty$ differ by an integer, the two Frobenius-series solutions become linearly dependent. However, due to the particular form of the coefficients in the differential equation, the second independent solution does not acquire a logarithmic term.} 

Combining \eqref{SchWHrelationbetweencoeff} and \eqref{SchWHkappa}, we obtain the leading-order expansion in $\omega$ of the real and imaginary parts of the Love numbers \eqref{SchWHLove}, keeping terms up to first order in $\epsilon$
\begin{equation}
    \begin{aligned}
        \Re k_{\text{Love}}
        &=\dfrac{(l!)^4}{2(2l)!(2l+1)!}\dfrac{1}{\ln\epsilon+2H_l}\left[1+\epsilon\left((l(l+1)+1)\left(\ln\epsilon+2H_l\right)+\dfrac{1+2l(l+1)}{\ln\epsilon+2H_l}\right)\right]+\mathcal{O}\left((\omega r_s)^2\right)
        \\
        \Im k_{\text{Love}}
        &=\dfrac{1}{8}\left(2\omega r_s\right)^{2l+1}\dfrac{(l!)^{10}}{\left[(2l)!(2l+1)!\right]^3}\dfrac{1}{\left(\ln\epsilon+2H_l\right)^2}\Bigg[1-2\epsilon\Bigg((l(l+1)+1)\left(\ln\epsilon+2H_l\right)-\dfrac{1+2l(l+1)}{\ln\epsilon+2H_l}\Bigg)\Bigg]+
        \\
        &+\mathcal{O}\left((\omega r_s)^{2l+3}\right),
    \end{aligned}
    \label{SchWHLoveReIm}
\end{equation}
where $H_l=\sum_{k=1}^{l}\frac{1}{k}$ is the harmonic number. Here we used the fact that, for super-throat modes $\omega r_s\ll\omega r_s\ln\frac{1}{\epsilon}\ll 1$, so that the factors $e^{\pm 2i\omega r_s\ln\epsilon}$ can be expanded in powers of $\omega r_s\ln\epsilon$. The result shows that the leading contribution in $\epsilon$ scales as $\frac{1}{\ln\epsilon+2H_l}$, in agreement with previous calculations \cite{Cardoso:2017cfl} of the static Love numbers for thin-shell wormholes. The leading contribution to the imaginary part scales as $\frac{1}{\left(\ln\epsilon+2H_l\right)^2}$.

The calculation confirms our earlier expectation \eqref{superthroatimlove} that the imaginary part of the Love numbers \eqref{SchWHLoveReIm} appears only at order $\omega^{2l+1}$. Using \eqref{imloveandtranscoef}, we can obtain the transmission coefficient
\begin{equation}
    \begin{aligned}
        \Gamma_l(\omega)
        &=\dfrac{1}{4}\left(2\omega r_s\right)^{4l+2}\left[\dfrac{(l!)^{3}}{(2l)!(2l+1)!}\right]^{4}\dfrac{1}{\left(\ln\epsilon+2H_l\right)^2}\Bigg\{1-2\epsilon\Bigg((l(l+1)+1)\left(\ln\epsilon+2H_l\right)-\dfrac{1+2l(l+1)}{\ln\epsilon+2H_l}\Bigg)\Bigg\}+
        \\
        &+\mathcal{O}\left((\omega r_s)^{4l+4}\right).
    \end{aligned}
\end{equation}

For $l=0$, the absorption cross section $\sigma_0=\frac{\pi}{\omega^2}\Gamma_0(\omega)$ should be independent of $\omega$ and should be determined only by an effective absorption area
\begin{equation}
    \sigma_0=\pi\dfrac{r_s^2}{\ln^2\epsilon}\left\{1-2\epsilon\left(\ln\epsilon-\dfrac{1}{\ln\epsilon}\right)\right\},
\end{equation}
which goes to zero as $\epsilon\rightarrow 0$. This means that the effective size of the wormhole decreases, so that the super-throat modes gradually become irrelevant, leaving only the sub-throat modes.

\begin{figure}[htbp]
    \centering
    \includegraphics[width=0.7\textwidth]{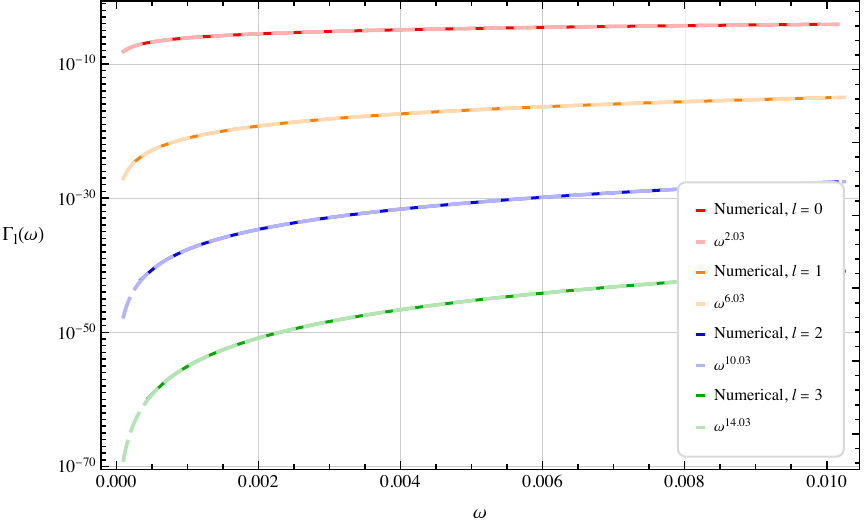}
    \caption{Numerically obtained transmission coefficient for super-throat modes of the thin-shell Schwarzschild wormhole (solid lines), with $\epsilon=10^{-5}$. The dashed lines show power-law fits to the numerical data.}
    \label{fig:SchWHtransmissioncoeffsuperthroat}
\end{figure}

We also numerically investigated the low-frequency behavior of the transmission coefficient and confirmed the $\Gamma_l(\omega)\sim\omega^{4l+2}$ behavior (see Fig.~\ref{fig:SchWHtransmissioncoeffsuperthroat}).

\subsection{The sub-throat modes}\label{sec:SchWHsubthroatmodes}

We now turn to calculating the Love numbers for the sub-throat modes. Recall that these modes also belong to the low-frequency regime, and therefore the near zone equation \eqref{SchWHscalareqnnearzone} remains valid. The difference from the super-throat modes lies in the boundary conditions imposed on the solution. As discussed in section \ref{sec:tworegims}, for these modes the throat effectively behaves as a horizon. From a dynamical point of view, an observer who can probe such modes does not wait long enough to receive the echo signal and therefore does not detect any signal coming from the left universe. For such an observer, the wormhole effectively appears as a black hole. 
Accordingly, we impose a purely ingoing boundary condition at the throat, or equivalently, require that no outgoing waves emerge from it.

To impose this condition, we first need to introduce a basis of waves at the throat. The general solution of the equation \eqref{SchWHscalareqnnearzone} can be written in a form analogous to \eqref{SchWHnearzonesolution}
\begin{equation}
    \begin{aligned}
        R(r)=\left(\dfrac{r}{r_s}\right)^l
        &\left[A_1\left(1-\dfrac{r_s}{r}\right)^{-i\omega r_s}{}_2F_1\left(-l,-l-2i\omega r_s,1-2i\omega r_s;1-\dfrac{r_s}{r}\right)+\right.
        \\
        &\left.+A_2\left(1-\dfrac{r_s}{r}\right)^{i\omega r_s}{}_2F_1\left(-l,-l+2i\omega r_s,1+2i\omega r_s;1-\dfrac{r_s}{r}\right)\right],
    \end{aligned}
    \label{SchWHsubthroatsol}
\end{equation}
where we omit the universe labels R and L, since we are interested only in the right universe.

At first sight, this solution may already appear to be written in the required basis of waves. However, since the throat is slightly shifted away from the horizon, the functions multiplying $A_1$ and $A_2$ do not describe purely ingoing and outgoing waves at the throat, respectively. Therefore, simply setting $A_2=0$ is not sufficient. 

Inside the throat, the effective potential for the scalar field \eqref{SchWHeffpot} develops a plateau (see Fig.~\ref{fig:effpot} for the analogous DS case) of length $r_s\ln\frac{1}{\epsilon}$, with a nonzero height $V_{l}(r_0)=\frac{\epsilon}{r^2_s}\left(l(l+1)+1\right)$. Consequently, propagating waves exist inside the throat only when $\omega^2>V_{l}(r_0)$. In this case, near the throat the solution can be written as
\begin{equation}
    \Psi(z_{\ast}\sim 0)\sim C_1e^{-i\frac{\chi}{r_s}z_{\ast}}+C_2e^{i\frac{\chi}{r_s}z_{\ast}},
\end{equation}
where $\chi^2=\omega^2r_s^2-\epsilon\left(l(l+1)+1\right)$. The general solution of \eqref{SchWHeffpot} can be expressed in terms of the basis of ingoing and outgoing waves at the throat (see appendix \ref{app:SCHwavebasis})
\begin{equation}
    \Psi(z_{\ast})=r(z_{\ast})\left[C_1W_{\text{in}}(z_{\ast})+C_2W_{\text{out}}(z_{\ast})\right].
\end{equation}
Imposing the absence of outgoing waves gives the following solution in terms of $R(r)$
\begin{equation}
    \begin{aligned}
        R(r)=\left(\dfrac{r}{r_s}\right)^l
        &\left[\left(1-\dfrac{r_s}{r}\right)^{-i\omega r_s}{}_2F_1\left(-l,-l-2i\omega r_s,1-2i\omega r_s;1-\dfrac{r_s}{r}\right)+\right.
        \\
        &\left.+\alpha_{\text{in}}\left(1-\dfrac{r_s}{r}\right)^{i\omega r_s}{}_2F_1\left(-l,-l+2i\omega r_s,1+2i\omega r_s;1-\dfrac{r_s}{r}\right)\right].
    \end{aligned}
\end{equation}
It is convenient to relate this solution to the general solution \eqref{SchWHsubthroatsol}. This gives the following relation between the coefficients $A_1$ and $A_2$
\begin{equation}
    \dfrac{A_2}{A_1}=-\left(\dfrac{\epsilon}{1+\epsilon}\right)^{-2i\omega r_s}\dfrac{(1+\epsilon)\left(\chi(1+\epsilon)^2-\omega r_s-i(l+1)\epsilon\right)H_{-}\left(\frac{\epsilon}{1+\epsilon}\right)-i\epsilon H'_{-}\left(\frac{\epsilon}{1+\epsilon}\right)}{(1+\epsilon)\left(\chi(1+\epsilon)^2+\omega r_s-i(l+1)\epsilon\right)H_{+}\left(\frac{\epsilon}{1+\epsilon}\right)-i\epsilon H'_{+}\left(\frac{\epsilon}{1+\epsilon}\right)},
    \label{SchWHsubthroatratio}
\end{equation}
where $H_{\pm}(x)={}_2F_1\left(-l,-l\pm 2i\omega r_s,1\pm 2i\omega r_s;x\right)$.

We can now use the expansion of the solution in the intermediate zone, where the Love numbers are defined,
\begin{equation}
    \begin{aligned}
        R(r)
        &\sim\left(1-\dfrac{r_s}{r}\right)^{-i\omega r_s}\left(\dfrac{r}{r_s}\right)^l\Bigg[Z^{\text{source}}(r)+k_{\text{Love}}\left(\dfrac{r_s}{r}\right)^{2l+1}Z^{\text{response}}(r)\Bigg],
    \end{aligned}
\end{equation}
with the source function $Z^{\text{source}}(r)$ and response function $Z^{\text{response}}(r)$ defined in Eq.~\eqref{SchWHsourceresposesplit}. The coefficients $k_{\text{Love}}$ are the Love numbers
\begin{equation}
    k_{\text{Love}}=\dfrac{\Gamma(-2l-1)\Gamma(l+1)\Gamma(l+1-2i\omega r_s)}{\Gamma(2l+1)\Gamma(-l)\Gamma(-l-2i\omega r_s)}\dfrac{1-\frac{A_2}{A_1}\frac{\Gamma(1+2i\omega r_s)\Gamma(l+1-2i\omega r_s)}{\Gamma(1-2i\omega r_s)\Gamma(l+1+2i\omega r_s)}}{1+\frac{A_2}{A_1}\frac{\Gamma(1+2i\omega r_s)\Gamma(l+1-2i\omega r_s)}{\Gamma(1-2i\omega r_s)\Gamma(l+1+2i\omega r_s)}},
    \label{SchWHsubthroatalmostLove}
\end{equation}
into which the ratio \eqref{SchWHsubthroatratio} should be substituted. 

Recall that throughout the intermediate steps we treated $l$ as a non-integer in order to avoid possible divergences. We now return to positive integer values of $l$. Fortunately, no divergences arise, since the poles of the two gamma functions $\Gamma(-2l-1)$ and $\Gamma(-l)$ cancel each other. 

Since the system exhibits a threshold such that propagating waves exist in the throat only when $\chi=\sqrt{\omega^2r_s^2-\epsilon\left(l(l+1)+1\right)}$ is real, while no propagating waves are present when $\chi$ is imaginary, we expect the imaginary part of the Love numbers to vanish as $\chi$ becomes imaginary. To verify this, we first note that, for integer $l$, the prefactor in \eqref{SchWHsubthroatalmostLove} involving the ratio of gamma functions is purely imaginary $\frac{\Gamma(-2l-1)\Gamma(l+1)\Gamma(l+1-2i\omega r_s)}{\Gamma(2l+1)\Gamma(-l)\Gamma(-l-2i\omega r_s)}=iS(\omega r_s)$, while the ratio of gamma functions in the numerator and denominator has unit modulus $\left|\frac{\Gamma(1+2i\omega r_s)\Gamma(l+1-2i\omega r_s)}{\Gamma(1-2i\omega r_s)\Gamma(l+1+2i\omega r_s)}\right|=1$. Therefore, the imaginary part of the Love numbers, when nonzero, takes the form
\begin{equation}
    \Im k_{\text{Love}}=S(\omega r_s)\dfrac{1-\left|\frac{A_2}{A_1}\right|^2}{\left|1+\frac{A_2}{A_1}\frac{\Gamma(1+2i\omega r_s)\Gamma(l+1-2i\omega r_s)}{\Gamma(1-2i\omega r_s)\Gamma(l+1+2i\omega r_s)}\right|^2}.
\end{equation}
The imaginary part is nonzero when $\left|\frac{A_2}{A_1}\right|\neq 1$ and vanishes otherwise. Since $H^{\ast}_{-}(x)=H_{+}(x)$ in \eqref{SchWHsubthroatratio}, we find that 
\begin{equation}
    \left|\dfrac{A_2}{A_1}\right|=\left|\dfrac{(1+\epsilon)\left(\chi(1+\epsilon)^2-\omega r_s-i(l+1)\epsilon\right)H_{-}\left(\frac{\epsilon}{1+\epsilon}\right)-i\epsilon H'_{-}\left(\frac{\epsilon}{1+\epsilon}\right)}{(1+\epsilon)\left(\chi(1+\epsilon)^2+\omega r_s-i(l+1)\epsilon\right)H_{+}\left(\frac{\epsilon}{1+\epsilon}\right)-i\epsilon H'_{+}\left(\frac{\epsilon}{1+\epsilon}\right)}\right|=1
\end{equation}
for imaginary $\chi$. Hence, the imaginary part of the Love numbers vanishes, in agreement with our expectations.

We emphasize, however, that the sub-throat modes considered in this section, by definition, satisfy $\omega r_s\ll 1\ll \omega r_s\ln\frac{1}{\epsilon}$ and lie in the regime where waves propagate through the throat and are well separated from the threshold $\left(\omega r_s\right)^2=\epsilon\left(l(l+1)+1\right)$. Therefore, the following expansion in $\epsilon$ and $\frac{\epsilon}{\omega r_s}$ is valid, since 
\begin{equation}
    \dfrac{\epsilon}{\omega r_s}\ll\epsilon\ln\dfrac{1}{\epsilon}\ll 1.
\end{equation}
We can therefore separate the Love numbers into their real and imaginary parts and expand them to leading order in $\omega r_s$
\begin{equation}
    \begin{aligned}
        \Re k_{\text{Love}}
        &=\epsilon\dfrac{(l!)^4\left(l(l+1)+1\right)}{(2l)!(2l+1)!}\Big[\cos\left(2\omega r_s\left(\ln\epsilon+2H_l\right)\right)-\dfrac{1}{2\omega r_s}\sin\left(2\omega r_s\left(\ln\epsilon+2H_l\right)\right)\Big]
        \\
        \Im k_{\text{Love}}
        &=\omega r_s\dfrac{(l!)^4}{(2l)!(2l+1)!}\Bigg[1-\epsilon\dfrac{\left(l(l+1)+1\right)}{2(\omega r_s)^2}\left(\cos\left(2\omega r_s\left(\ln\epsilon+2H_l\right)\right)+2\omega r_s\sin\left(2\omega r_s\left(\ln\epsilon+2H_l\right)\right)\right)\Bigg].
    \end{aligned}
    \label{SchWHsubthroatLovenumbers}
\end{equation}
It is important to note that, for the sub-throat modes, $1\ll\omega r_s\ln\frac{1}{\epsilon}$ and therefore the trigonometric functions cannot be expanded in their arguments. We keep them in their exact form. However, all trigonometric terms appear with a prefactor proportional to $\epsilon$. Since the sine and cosine functions are bounded, these terms have a well-defined limit as $\epsilon\rightarrow 0$. 



\section{Damour-Solodukhin wormhole}\label{sec:DSWH}
\subsection{Wormhole geometry}\label{sec:DSWHgeometry}

In this section, we consider the Damour-Solodukhin wormhole \cite{Damour:2007ap}, which is described by the metric 
\begin{equation}
    \text{d}s^2=-\left(1-\dfrac{r_s}{r}+\lambda^2\right)\text{d}t^2+\dfrac{\text{d}r^2}{1-\frac{r_s}{r}}+r^2\left(\text{d}\theta^2+\sin^2\theta\text{d}\phi\right).
\end{equation}
For our purposes, it is more convenient to rescale the time coordinate $t\rightarrow \frac{t}{\sqrt{1+\lambda^2}}$ and introduce the parameter $\eta=\frac{\lambda^2}{1+\lambda^2}$. The metric then takes the form
\begin{equation}
    \text{d}s^2=-\left(1-\dfrac{r_s(1-\eta)}{r}\right)\text{d}t^2+\dfrac{\text{d}r^2}{1-\frac{r_s}{r}}+r^2\left(\text{d}\theta^2+\sin^2\theta\text{d}\phi\right).
    \label{DSWHmetric}
\end{equation}
With this parametrization, the throat is located at $r_{\text{th}}=r_s$. Unlike the thin-shell Schwarzschild wormhole geometry (see section  \ref{sec:schwormholegeometry}), the DS geometry is smooth across the throat. 

Within GR, this geometry is supported by an effective matter distribution with the following nonvanishing components of the stress-energy tensor $T^{\mu}_{\nu}=\text{diag}\left(-\rho,p_r,p_t,p_t\right)$
\begin{equation}
    \rho=0,\quad p_r=-\dfrac{1}{8\pi r^2}\dfrac{r_s\eta}{r-r_s(1-\eta)},\quad p_t=\dfrac{1}{32\pi r^2}\dfrac{r_s\eta\left(2r-r_s(1-\eta)\right)}{\left(r-r_s(1-\eta)\right)^2}.
\end{equation}
Thus, the DS wormhole violates both the null and weak energy conditions. 

The tortoise coordinate corresponding to this metric is given by 
\begin{equation}
    \dfrac{\text{d}z_{\ast}}{\text{d}r}=\pm\dfrac{r}{\sqrt{\left(r-r_s\right)\left(r-r_s(1-\eta)\right)}},
    \label{DSWHtortoisecoordinate}
\end{equation}
where the $+$ sign corresponds to the right universe and the $-$ sign to the left universe. Without loss of generality, we choose $z_{\ast}(r_s)=0$. Near the throat, the tortoise coordinate behaves as 
\begin{equation}
    z_{\ast}\approx \pm 2\sqrt{\dfrac{r_s}{\eta}}\sqrt{r-r_s}\quad\Rightarrow\quad r-r_s\approx \eta\dfrac{z^2_{\ast}}{4r_s},
    \label{DSWHtortoisecoordinatenearthroat}
\end{equation}
confirming that the geometry is smooth across the throat.

\subsection{Scalar field equation}\label{sec:scalarfieldequationDSWH}

For the DS wormhole $\Delta(r)=r(r-r_s)$ and $\bar{\Delta}(r)=r(r-r_s(1-\eta))$. Consequently, the scalar field equation takes the form \eqref{scalreq}
\begin{equation}
    \dfrac{\text{d}^2}{\text{d}r^2}R(r)+\left(\dfrac{1}{r}+\dfrac{1}{2(r-r_s)}+\dfrac{1}{2(r-r_s(1-\eta))}\right)\dfrac{\text{d}}{\text{d}r}R(r)+\dfrac{\omega^2r^4-l(l+1)r(r-r_s(1-\eta))}{r^2(r-r_s)(r-r_s(1-\eta))}R(r)=0.
    \label{DSWHscalarequation}
\end{equation}
Introducing the function $\Psi=r(z_{\ast})R(z_{\ast})$ and rewriting the equation in terms of the tortoise coordinate \eqref{DSWHtortoisecoordinate}, we obtain a Schrödinger-like equation with an effective potential
\begin{equation}
    \dfrac{\text{d}^2}{\text{d}z_{\ast}^2}\Psi+\left(\omega^2-V_l\right)\Psi=0,\quad V_l(r)=r_s\dfrac{r(2-\eta)-2r_s(1-\eta)}{2r^4}+\left(1-\dfrac{r_s(1-\eta)}{r}\right)\dfrac{l(l+1)}{r^2}.
    \label{DSWHeffectivepotential}
\end{equation}
It is important to note that $V_l(r_s)=\frac{\eta}{2r_s^2}\left(1+2l(l+1)\right)$; therefore, the effective potential does not vanish at the throat. As a result, waves propagate inside the throat only when $\omega^2>V_l(r_s)$. This property will play an important role in the analysis of the sub-throat modes. 

To obtain the near zone equation, we perform the modification $\omega^2r^4\rightarrow\omega^2r_s^4$ in \eqref{DSWHscalarequation} as described in section  \ref{sec:nearfarapp}. The resulting equation
has four regular singular points located at $r=0$, $r=r_s$, $r=r_s(1-\eta)$ and $r=\infty$, and therefore can be reduced to the Heun equation. To this end, we first introduce the dimensionless coordinate $z=\frac{r_s-r}{r_s}$, under which the singular points are mapped to $z=1$, $z=0$, $z=\eta$ and $z=\infty$, respectively. The equation \eqref{DSWHscalarequation} takes the form
\begin{equation}
    \dfrac{\text{d}^2}{\text{d}z^2}R(z)+\left(\dfrac{1}{2z}+\dfrac{1}{z-1}+\dfrac{1}{2(z-\eta)}\right)\dfrac{\text{d}}{\text{d}z}R(z)+\dfrac{\omega^2r_s^2-l(l+1)(z-1)(z-\eta)}{z(z-1)^2(z-\eta)}R(z)=0.
\end{equation}
Near the point $z=1$, the solution behaves as $R(z)\sim (z-1)^{\pm i p}$, where $p=\frac{\omega r_s}{\sqrt{1-\eta}}\approx\omega r_s\left(1+\frac{\eta}{2}\right)$. Since the point $z=1$, corresponding to $r=r_s(1-\eta)$, lies outside the physical region ($r\geq r_s$), the choice of exponent is irrelevant.\footnote{The choice of sign does not affect the connection problem considered below, which relates the solutions at $z=0$ and $z=\infty$. As a consequence, all connection coefficients will be even functions of $p$, and therefore of $\omega r_s$ (see appendix \ref{app:connection}).} For convenience, we choose the positive sign. We can therefore introduce $R(z)=(1-z)^{ip}u(z)$, and the differential equation takes the standard Heun form
\begin{equation}
    \begin{aligned}
        \dfrac{\text{d}^2}{\text{d}z^2}u(z)+\Bigg(\dfrac{1}{2z}+
        &\dfrac{1+2ip}{z-1}+\dfrac{1}{2(z-\eta)}\Bigg)\dfrac{\text{d}}{\text{d}z}u(z)+
        \\
        &+\dfrac{(-l+ip)(l+1+ip)z-\left(-\eta l(l+1)+\eta\frac{ip}{2}+p^2(1-\eta)\right)}{z(z-1)(z-\eta)}u(z)=0,
    \end{aligned}
    \label{DSWHeqnearZone}
\end{equation}
with the parameters (see appendix \ref{app:HeunEq} for details on the Heun equation)
\begin{equation}
    \alpha=-l+ip,\quad\beta=l+1+ip,\quad\gamma=\dfrac{1}{2},\quad\delta=1+2ip,\quad\varepsilon=\dfrac{1}{2},\quad q=-\eta l(l+1)+\eta\frac{ip}{2}+p^2(1-\eta).
\end{equation}

In what follows, we write the basis solutions near the two singularities of interest: $z=0$, which corresponds to $r=r_s$, and $z=\infty$, which corresponds to $r=\infty$. In the near zone approximation, the latter actually corresponds to the intermediate zone.

Solutions near $z=0$:
\begin{equation}
    \begin{aligned}
        &u^{(0)}_{-}(z)=\text{HeunG}\left(\eta,q\left|-l+ip,l+1+ip,\dfrac{1}{2},1+2ip\right|z\right)
        \\
        &u^{(0)}_{+}(z)=z^{1/2}\text{HeunG}\left(\eta,q+\frac{\eta}{2}(1+2ip)+\frac{1}{4}\left|-l+\dfrac{1}{2}+ip,l+\dfrac{3}{2}+ip,\dfrac{3}{2},1+2ip\right|z\right)
    \end{aligned}
    \label{DSWHsolutionsnearz0}
\end{equation}

Solutions near $z=\infty$:
\begin{equation}
    \begin{aligned}
        &u^{(\infty)}_{+}(z) = z^{l-ip}\text{HeunG}\left(\eta,l^2-\dfrac{l}{2}(\eta+4ip)\left|-l+ip,-l+ip+\dfrac{1}{2},-2l,\dfrac{1}{2}\right|\dfrac{\eta}{z}\right)
        \\
        &u^{(\infty)}_{-}(z) = z^{-l-1-ip}\text{HeunG}\left(\eta,(l+1)^2+\dfrac{l+1}{2}(\eta+4ip)\left|l+1+ip,l+\dfrac{3}{2}+ip,2l+2,\dfrac{1}{2}\right|\dfrac{\eta}{z}\right)
    \end{aligned}
    \label{DSWHsolutionsnearzinfty}
\end{equation}
Recall that we use a prescription in which $l$ is treated as a non-integer number throughout the calculation. Only at the end of the calculation do we restore positive integer values of $l$ by making the shift $l\rightarrow l-\delta$, which allows us to isolate possible divergences in $\delta$. This prescription avoids potential problems associated with the parameter $-2l$ in the function $u^{(\infty)}_{+}(z)$ (see appendix \ref{app:HeunEq} for details). Then the general solutions of \eqref{DSWHeqnearZone} in both universes have the following form
\begin{equation}
    R^{\text{(R,L)}}(z) = (1-z)^{ip}\left[A^{\text{(R,L)}}_1 u^{(0)}_{-}(z)+A^{\text{(R,L)}}_2 u^{(0)}_{+}(z)\right]
    \label{DSWHgeneralSolution}
\end{equation}

We now need to relate the solutions near $z=0$ to those at $z=\infty$, corresponding to the intermediate zone. To do so, we use the connection formulas derived recently in \cite{Bonelli:2022ten}, which follow from the semiclassical approximation to solutions of the BPZ equations together with the AGT correspondence. A perturbative derivation of these formulas can be found in \cite{Lisovyy:2022flm}. 
\begin{equation}
    \begin{aligned}
        &u^{(0)}_{-}(z) = (-1)^{l-ip} C^{(0)+}_{-(\infty)}u^{(\infty)}_{+}(z)+(-1)^{-l-1-ip} C^{(0)-}_{-(\infty)}u^{(\infty)}_{-}(z)
        \\
        &u^{(0)}_{+}(z) = -i(-1)^{l-ip} C^{(0)+}_{+(\infty)}u^{(\infty)}_{+}(z)-i(-1)^{-l-1-ip} C^{(0)-}_{+(\infty)}u^{(\infty)}_{-}(z).
        \label{DSWHconection}
    \end{aligned}
\end{equation}
The additional multiplicative factors $(-1)^{l-ip}$, $(-1)^{-l-1-ip}$ and $-i$ arise because all singular points of the differential equation \eqref{DSWHeqnearZone} lie on the positive real axis $z\geq 0$. We therefore connect the basis solutions at the two points of interest by analytically continuing them through the lower half-plane $\arg z=-\pi$. 

Returning to the original radial coordinate $r$, we obtain the following
\begin{equation}
    \begin{aligned}
        &u^{(0)}_{-}(r)=\left(1-\dfrac{r_s}{r}\right)^{-ip}\left[C^{(0)+}_{-(\infty)}\left(\dfrac{r}{r_s}\right)^{l-ip}Z^{\text{source}}(r)+C^{(0)-}_{-(\infty)}\left(\dfrac{r}{r_s}\right)^{-l-1-ip}Z^{\text{response}}(r)\right]
        \\
        &u^{(0)}_{+}(r)=-i\left(1-\dfrac{r_s}{r}\right)^{-ip}\left[C^{(0)+}_{+(\infty)}\left(\dfrac{r}{r_s}\right)^{l-ip}Z^{\text{source}}(r)+C^{(0)-}_{+(\infty)}\left(\dfrac{r}{r_s}\right)^{-l-1-ip}Z^{\text{response}}(r)\right].
    \end{aligned}
\end{equation}
Here we have introduced the source function, which describes the external tidal field including all gravitational corrections, and the response function, which describes the response of the system to the external perturbation
\begin{equation}
    \begin{aligned}
        &Z^{\text{source}}(r) = \left(1-\dfrac{r_s}{r}\right)^{l}\text{HeunG}\left(\eta,l^2-\dfrac{l}{2}(\eta+4ip)\left|-l+ip,-l+\dfrac{1}{2}+ip,-2l,\dfrac{1}{2}\right|\dfrac{\eta r_s}{r_s-r}\right)
       \\
        &Z^{\text{response}}(r) = \left(1-\dfrac{r_s}{r}\right)^{-l-1}\text{HeunG}\left(\eta,(l+1)^2+\dfrac{l+1}{2}(\eta+4ip)\left|l+1+ip,l+\dfrac{3}{2}+ip,2l+2,\dfrac{1}{2}\right|\dfrac{\eta r_s}{r_s-r}\right).
    \end{aligned}
    \label{sourceresponsesplitdynDS}
\end{equation}
Both functions approach unity as $r\rightarrow \infty$, and they are mapped into one another under $l\leftrightarrow -l-1$. The solution in the intermediate zones in both universes takes the form 
\begin{equation}
    \begin{aligned}
        R^{\text{(R,L)}}(r)=\left(1-\dfrac{r_s}{r}\right)^{-ip}\Bigg[
        &\left(A^{\text{(R,L)}}_1C^{(0)+}_{-(\infty)}-A^{\text{(R,L)}}_2iC^{(0)+}_{+(\infty)}\right)\left(\frac{r}{r_s}\right)^lZ^{\text{source}}(r)+
        \\
        &+\left(A^{\text{(R,L)}}_1C^{(0)-}_{-(\infty)}-A^{\text{(R,L)}}_2iC^{(0)-}_{+(\infty)}\right)\left(\frac{r}{r_s}\right)^{-l-1}Z^{\text{response}}(r)\Bigg]
    \end{aligned}
\end{equation}
\subsection{The super-throat modes}\label{sec:thirdregime}

We begin by computing the Love numbers for the super-throat modes. Since both asymptotically flat universes are relevant for these modes, the solutions in the right and left universes must be matched across the throat. Because the effective potential \eqref{DSWHeffectivepotential} is continuous across the throat and contains no terms proportional to a delta function, the scalar field satisfies the following matching conditions at the throat
\begin{equation}
    \begin{cases}
        \Psi^{\text{(R)}}(z_{\ast}=0) = \Psi^{\text{(L)}}(z_{\ast}=0)
        \vspace{0.2cm}
        \\
        \dfrac{\text{d}}{\text{d}z_{\ast}}\Psi^{\text{(R)}}(z_{\ast}=0) = \dfrac{\text{d}}{\text{d}z_{\ast}}\Psi^{\text{(L)}}(z_{\ast}=0)
    \end{cases},
\end{equation}
where $\Psi^{\text{(R)}}$ and $\Psi^{\text{(L)}}$ denote the solution in the right and left universes, respectively. These matching conditions are equivalent to requiring that the scalar field flux $\mathcal{F}=\frac{1}{2i}\left(\Psi^{\dagger}\partial_{z_{\ast}}\Psi-\Psi\partial_{z_{\ast}}\Psi^{\dagger}\right)$ is conserved across the throat: $\mathcal{F}^{(\text{R})}=\mathcal{F}^{(\text{L})}$. Using the explicit form of the solutions \eqref{DSWHgeneralSolution}, one obtains the following relations between the coefficients in the two universes: $A^{\text{(L)}}_1=A^{\text{(R)}}_1=A_1$ and $A^{\text{(L)}}_2=-A^{\text{(R)}}_2=-A_2$. These relations can also be understood from the fact that the wormhole throat is smooth. Therefore, the solutions in the two universes must join smoothly when expressed in a global coordinate system, for instance, the tortoise coordinate \eqref{DSWHtortoisecoordinate}. The solution $u^{(0)}_{+}(z)$ \eqref{DSWHsolutionsnearz0} contains only half-integer powers of $z$, which become odd powers in the tortoise coordinate \eqref{DSWHtortoisecoordinatenearthroat}, whereas $u^{(0)}_{-}(z_{\ast})$ contains even powers. Consequently, the coefficients in \eqref{DSWHgeneralSolution} multiplying $u^{(0)}_{-}(z)$ must be the same in the two universes, while those multiplying $u^{(0)}_{+}(z)$ must have opposite signs.

We must now impose the boundary condition at the left asymptotic infinity. Recall that the source of the external tidal field is assumed to be localized in the right universe. Therefore, waves transmitted into the left universe are not reflected back, and the condition \eqref{conditioninleftuniversedyn} must be imposed there. To implement this condition, we first match the near zone solution in the left universe
\begin{equation}
    \begin{aligned}
        R^{\text{(L)}}(r)=\left(1-\dfrac{r_s}{r}\right)^{-ip}\Bigg[
        &\left(A_1C^{(0)+}_{-(\infty)}+A_2iC^{(0)+}_{+(\infty)}\right)\left(\frac{r}{r_s}\right)^lZ^{\text{source}}(r)+
        \\
        &+\left(A_1C^{(0)-}_{-(\infty)}+A_2iC^{(0)-}_{+(\infty)}\right)\left(\frac{r}{r_s}\right)^{-l-1}Z^{\text{response}}(r)\Bigg]
    \end{aligned}
\end{equation}
to the far zone solution \eqref{farzoneintermatch} in the intermediate zone of the left universe, and then impose the condition \eqref{conditioninleftuniversedyn}. This yields the following relation between the coefficients
\begin{equation}
    \dfrac{A_1}{A_2}=-i\dfrac{C^{(0)+}_{+(\infty)}}{C^{(0)+}_{-(\infty)}}\underbrace{\dfrac{1-i\left(\omega r_s\right)^{2l+1}\frac{C^{(0)-}_{+(\infty)}}{C^{(0)+}_{+(\infty)}}\frac{2^{2l}\Gamma^2(l+1)}{\Gamma(2l+1)\Gamma(2l+2)}}{1-i\left(\omega r_s\right)^{2l+1}\frac{C^{(0)-}_{-(\infty)}}{C^{(0)+}_{-(\infty)}}\frac{2^{2l}\Gamma^2(l+1)}{\Gamma(2l+1)\Gamma(2l+2)}}}_{\kappa}.
    \label{DSWHcoefrelationkappa}
\end{equation}

The Love numbers are defined in the intermediate region of the right universe, where the solution takes the form
\begin{equation}
    \begin{aligned}
        R^{\text{(R)}}(r)=\left(1-\dfrac{r_s}{r}\right)^{-ip}\Bigg[
        &\left(A_1C^{(0)+}_{-(\infty)}-A_2iC^{(0)+}_{+(\infty)}\right)\left(\frac{r}{r_s}\right)^lZ^{\text{source}}(r)+
        \\
        &+\left(A_1C^{(0)-}_{-(\infty)}-A_2iC^{(0)-}_{+(\infty)}\right)\left(\frac{r}{r_s}\right)^{-l-1}Z^{\text{response}}(r)\Bigg].
    \end{aligned}
    \label{DSWHsolutioninfty}
\end{equation}
Accordingly, the Love numbers are given by the ratio of the coefficient multiplying the response function to the coefficient multiplying the source function
\begin{equation}
    k_{\text{Love}}=\dfrac{A_1C^{(0)-}_{-(\infty)}-A_2iC^{(0)-}_{+(\infty)}}{A_1C^{(0)+}_{-(\infty)}-A_2iC^{(0)+}_{+(\infty)}}=\dfrac{C^{(0)-}_{-(\infty)}}{C^{(0)+}_{-(\infty)}}\dfrac{\kappa}{1+\kappa}+\dfrac{C^{(0)-}_{+(\infty)}}{C^{(0)+}_{+(\infty)}}\dfrac{1}{1+\kappa}.
    \label{DSWHalmostLovenumbers}
\end{equation}

As in the thin-shell case discussed in section \ref{sec:SchWHsuperthroatmodes}, the leading contributions from the near and far zones enter at different orders in frequency. Therefore, despite working in the low-frequency approximation $\omega r_s\ll 1$, naively discarding all terms beyond linear order in $\omega r_s$ in the final expression \eqref{DSWHalmostLovenumbers} would be incorrect, as it would remove the leading nontrivial imaginary contribution. The correct procedure is to retain the leading contribution in $\omega r_s$ arising separately from both the near and far regions. In particular, one cannot simply set $\kappa=1$ \eqref{DSWHcoefrelationkappa} on the grounds that it contains higher powers of $\omega r_s$. The combination $(\omega r_s)^{2l+1}$, which appears in both the numerator and denominator of $\kappa$, represents the leading order contribution from the far zone \eqref{farzoneintermatch} and therefore cannot be neglected. Accordingly, $\kappa$ must be expanded while retaining its leading nontrivial dependence on $\omega r_s$
\begin{equation}
    \kappa=1-i\left(\omega r_s\right)^{2l+1}\frac{2^{2l}\Gamma^2(l+1)}{\Gamma(2l+1)\Gamma(2l+2)}\left(\frac{C^{(0)-}_{+(\infty)}}{C^{(0)+}_{+(\infty)}}-\frac{C^{(0)-}_{-(\infty)}}{C^{(0)+}_{-(\infty)}}\right)
    \label{DSWHkappaexp}
\end{equation}
The same reasoning applies to the connection coefficients $C^{(0)\pm}_{\pm(\infty)}$, which originate from the near zone solution. In this case, however, the leading contribution is independent of $\omega r_s$. Therefore, it is sufficient to retain only the first order term in their low-frequency expansion, but as we show in appendix \ref{app:connection}, the connection coefficients are even in $\omega r_s$.

Combining \eqref{DSWHalmostLovenumbers} and \eqref{DSWHkappaexp}, we obtain the following expression for the Love numbers, including the leading imaginary part 
\begin{equation}
    k_{\text{Love}}=\dfrac{1}{2}\left(\frac{C^{(0)-}_{+(\infty)}}{C^{(0)+}_{+(\infty)}}+\frac{C^{(0)-}_{-(\infty)}}{C^{(0)+}_{-(\infty)}}\right)+\dfrac{1}{8}i\left(2\omega r_s\right)^{2l+1}\dfrac{\Gamma^2(l+1)}{\Gamma(2l+1)\Gamma(2l+2)}\left(\frac{C^{(0)-}_{+(\infty)}}{C^{(0)+}_{+(\infty)}}-\frac{C^{(0)-}_{-(\infty)}}{C^{(0)+}_{-(\infty)}}\right)^2.
    \label{DSWHalmostLovenumbers2}
\end{equation}
We must now take  $l$ to positive integer values. To this end, we make the shift $l\rightarrow l-\delta$, which allows us to consistently account for all potential divergences in $\delta$. As can be seen from the explicit expressions for the connection coefficients (see appendix \ref{app:connection} for details), two of the coefficients $C^{(0)-}_{+(\infty)}$ and $C^{(0)-}_{-(\infty)}$ diverge; consequently, the first term in \eqref{DSWHalmostLovenumbers2} inherits this divergence, whereas in the second term the divergent contributions cancel. As a result, the Love numbers acquire the following structure
\begin{equation}
    k_{\text{Love}}=-\eta\dfrac{(l!)^2l(l+1)}{4(2l)!(2l+1)!}\prod_{k=1}^{l}\left(k^2+4p^2\right)\dfrac{1}{\delta}+k^{(\text{finite})}.
    \label{DSWHdivfinite}
\end{equation}

The complete solution \eqref{DSWHsolutioninfty}, however, is regular. The divergence of the Love numbers is exactly canceled by the divergence of the source function \eqref{sourceresponsesplitdynDS} (see appendix \ref{app:zsource} for details)
\begin{equation}
    Z^{\text{source}}(r)=\dfrac{1}{\delta}\eta\dfrac{(l!)^2l(l+1)}{4(2l)!(2l+1)!}\prod_{k=1}^{l}\left(k^2+4p^2\right)\left(\dfrac{r_s}{r}\right)^{2l+1}Z^{\text{response}}(r)+\mathcal{O}(\delta^0).
\end{equation}

From the EFT point of view \cite{Charalambous:2023jgq, Ivanov:2022hlo, Barbosa:2025uau, Cano:2025zyk, Ivanov:2024sds}, this divergence is identified with the divergence of the Wilson coefficients appearing in the finite-size effective action and is interpreted as the classical renormalization group (RG) flow of these coefficients. The residue of the divergence is precisely the $\beta$-function\footnote{The additional factor of $-2$ arises because the Love coefficient $k_{\text{Love}}$ is defined as a coefficient multiplying the power $\left(\frac{r_s}{r}\right)^{2l+1-2\delta}$.} 
\begin{equation}
    \beta=\mu\dfrac{\text{d}}{\text{d}\mu}k_{\text{Love}}=\eta\dfrac{(l!)^2l(l+1)}{2(2l)!(2l+1)!}\prod_{k=1}^{l}\left(k^2+4p^2\right),
    \label{DSWHbetafunction}
\end{equation}
where $\mu$ is the renormalization energy scale, or equivalently an inverse length scale. Thus, the Love numbers of the DS wormhole run and give rise to a logarithmic contribution
\begin{equation}
    k^{(\text{ren.})}_{\text{Love}}=\eta\dfrac{(l!)^2l(l+1)}{2(2l)!(2l+1)!}\prod_{k=1}^{l}\left(k^2+4p^2\right)\ln\dfrac{\mu}{\mu_0}+k^{(\text{finite})}.
\end{equation}
In contrast to the running part, which is universal, the finite part is ambiguous and therefore depends on the choice of the reference scale $\mu_0$ \cite{Ivanov:2022hlo}.

From the viewpoint of the theory of differential equations, the logarithmic contribution originates from the fact that, for positive integer values of $l$, the solutions near $z=\infty$ become degenerate: the characteristic exponents differ by an integer. Consequently, only one solution can be represented by a Frobenius series while the second necessarily contains a logarithmic term. 

In what follows, we choose $\mu=\frac{1}{r}$ and $\mu_0=\frac{1}{r_s}$, and retain the finite part in the final expression. For the super-throat modes $\omega r_s\ll\omega r_s\ln\frac{4}{\eta}\ll 1$, so the functions $\sin(px)$ and $\cos(px)$ appearing in the connection coefficients (see appendix \ref{app:connection}) can be expanded as power series in $px$ (recall $p=\frac{\omega r_s}{\sqrt{1-\eta}}$). Expanding all coefficients to leading order in $p$, while retaining the leading contribution to the imaginary part, we finally obtain the following expression for the Love numbers
\begin{equation}
    \begin{aligned}
        \Re k_{\text{Love}}&=
        \dfrac{(l!)^4}{(2l)!(2l+1)!}\Bigg\{-\eta\dfrac{l(l+1)}{2}\ln\dfrac{r}{r_s}+\dfrac{1}{2\left(\ln\frac{\eta}{4}+2H_l\right)}\left(1-\dfrac{\eta}{2}\dfrac{1+(2l+1)\left(\ln\frac{\eta}{4}+2H_l\right)}{\ln\frac{\eta}{4}+2H_l}\right)+
        \\
        &+\eta\dfrac{l(l+1)}{4}\Bigg[\dfrac{1}{3}\left(\ln\frac{\eta}{4}+2H_l\right)^2+\left(\ln\frac{\eta}{4}+2H_l\right)\left(\ln\frac{\eta}{4}+6H_l-2H_{2l+1}-2H_{2l}\right)-2H^{(2)}_l-\dfrac{\pi^2}{3}-
        \\
        &-\dfrac{2H^{(3)}_l+6\zeta(3)}{3\left(\ln\frac{\eta}{4}+2H_l\right)}\Bigg]\Bigg\}+\mathcal{O}\left((\omega r_s)^2\right)
        \\
        \Im k_{\text{Love}}&=\dfrac{1}{8}(2\omega r_s)^{2l+1}\dfrac{(l!)^{10}}{\left((2l)!(2l+1)!\right)^3}\dfrac{1}{\left(\ln\frac{\eta}{4}+2H_l\right)^2}\Bigg\{1-\eta\dfrac{1+(2l+1)\left(\ln\frac{\eta}{4}+2H_l\right)}{\left(\ln\frac{\eta}{4}+2H_l\right)}-
        \\
        &-\eta l(l+1)\Bigg[\dfrac{2}{3}\left(\ln\frac{\eta}{4}+2H_l\right)^2+2H^{(2)}_l+\dfrac{\pi^2}{3}+\dfrac{2H^{(3)}_l+6\zeta(3)}{3\left(\ln\frac{\eta}{4}+2H_l\right)}\Bigg]\Bigg\}+\mathcal{O}\left((\omega r_s)^{2l+3}\right),
    \end{aligned}
    \label{DSWHsuperthroatLovemodes}
\end{equation}
where $H^{(r)}_l=\sum_{k=1}^{l}\frac{1}{k^r}$ is a generalized harmonic number, $\zeta(z)$ is the Riemann zeta function, and $\eta=\frac{\lambda^2}{1+\lambda^2}$ is the deformation parameter introduced in Eq.~\eqref{DSWHmetric}. As in the thin-shell case discussed in section \ref{sec:SchWHsuperthroatmodes}, the leading contribution in $\eta$ scales as $\frac{1}{\ln\frac{\eta}{4}+2H_l}$, in agreement with previous calculations of the static Love numbers for wormholes \cite{Cardoso:2017cfl}. The leading contribution to the imaginary part again scales as $\frac{1}{\left(\ln\frac{\eta}{4}+2H_l\right)^2}$. However, there is an important difference between the Damour-Solodukhin and thin-shell cases: in the DS case, an additional logarithmic contribution, 
$\ln\dfrac{r}{r_s}$, appears in the real part of the Love numbers.  This suggests that a Wilsonian interpretation of the Love numbers is relevant in this case.

The calculation confirms our earlier expectation \eqref{superthroatimlove} that the imaginary part of the Love numbers for super-throat modes appears only at order $\omega^{2l+1}$. Using \eqref{imloveandtranscoef}, we obtain
\begin{equation}
    \begin{aligned}
        \Gamma_l(\omega)
        &=\dfrac{1}{4}(2\omega r_s)^{4l+2}\left[\dfrac{(l!)^{3}}{(2l)!(2l+1)!}\right]^4\dfrac{1}{\left(\ln\frac{\eta}{4}+2H_l\right)^2}\Bigg\{1-\eta\dfrac{1+(2l+1)\left(\ln\frac{\eta}{4}+2H_l\right)}{\left(\ln\frac{\eta}{4}+2H_l\right)}-
        \\
        &-\eta l(l+1)\Bigg[\dfrac{2}{3}\left(\ln\frac{\eta}{4}+2H_l\right)^2+2H^{(2)}_l+\dfrac{\pi^2}{3}+\dfrac{2H^{(3)}_l+6\zeta(3)}{3\left(\ln\frac{\eta}{4}+2H_l\right)}\Bigg]\Bigg\}+\mathcal{O}\left((\omega r_s)^{4l+4}\right).
    \end{aligned}
\end{equation}

For $l=0$, the absorption cross section $\sigma_0=\frac{\pi}{\omega^2}\Gamma_0(\omega)$ is independent of $\omega$ and is determined only by an effective absorption area
\begin{equation}
    \sigma_0=\dfrac{\pi r_s^2}{\ln^2\frac{\eta}{4}}\Bigg\{1-\eta\dfrac{1+\ln\frac{\eta}{4}}{\ln\frac{\eta}{4}}\Bigg\},
\end{equation}
which approaches zero logarithmically as $\eta\rightarrow 0$. This indicates that, in this limit, the effective size of the wormhole decreases, and the super-throat modes gradually become irrelevant. 

\begin{figure}[htbp]
    \centering
    \includegraphics[width=0.7\textwidth]{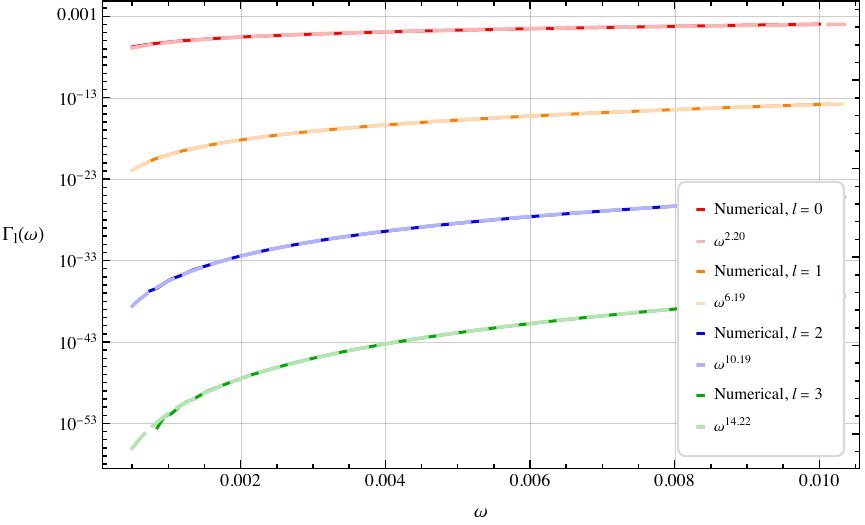}
    \caption{Numerically obtained transmission coefficient for super-throat modes of the DS wormhole (solid lines), with $\eta=10^{-10}$. The dashed lines show power-law fits to the numerical data.}
    \label{fig:DSWHtransmissioncoeffsuperthoat}
\end{figure}

We also studied the low-frequency behavior of the transmission coefficient numerically and confirmed the expected scaling $\Gamma_l(\omega)\sim \omega^{4l+2}$ (see Fig.~\ref{fig:DSWHtransmissioncoeffsuperthoat}).

\subsection{The sub-throat modes}\label{sec:DSWHsubthroat}

We now turn to the calculation of the Love numbers for the sub-throat modes. These modes also belong to the low-frequency regime, so the near zone equation \eqref{DSWHeqnearZone} remains valid. However, we must now impose a different boundary condition. As discussed in section \ref{sec:tworegims}, for these modes the throat effectively behaves as a horizon, since a distant observer does not wait long enough to receive the wave reflected from the left universe. Therefore, the appropriate boundary condition is the absence of an outgoing wave from the throat. 

The general solution of the equation \eqref{DSWHeqnearZone} can be written in a form analogous to \eqref{DSWHgeneralSolution} 
\begin{equation}
    R(z) = (1-z)^{ip}\left[A_1 u^{(0)}_{-}(z)+A_2 u^{(0)}_{+}(z)\right],
    \label{DSWHgeneralSolutionsubthroat}
\end{equation}
where we omit the R and L labels, since we are now interested only in the right universe. However, the solutions $u_{+}(z)$ and $u_{-}(z)$ do not form a wave basis near the throat. Due to the behavior of the tortoise coordinate near the throat \eqref{DSWHtortoisecoordinatenearthroat} (recall that $r-r_s=-r_s z$), $u_{-}(z_{\ast})$ and $u_{+}(z_{\ast})$ correspond to even and odd solutions in the tortoise coordinate, respectively. 

Near the throat, the effective potential develops a long plateau of length $r_s\ln\frac{4}{\eta}$ (see Fig.~\ref{fig:effpot}), with a nonzero height determined by the value of the potential at the throat $V_l(r_s)=\frac{\eta}{2r^2_s}(1+2l(l+1))$. Therefore, waves can propagate along the throat only if $\omega^2>V_l(r_s)$. 

Near the throat, the solution of \eqref{DSWHeffectivepotential} can thus be written as
\begin{equation}
    \Psi(z_{\ast}\sim 0)\sim C_1e^{-i\frac{\chi}{r_s}z_{\ast}}+C_2e^{i\frac{\chi}{r_s}z_{\ast}},
\end{equation}
where $\chi^2=\omega^2r_s^2-\frac{\eta}{2}(1+2l(l+1))$. We now transform from the even/odd basis to the basis of ingoing and outgoing waves $W_{\text{in/out}}$ (see appendix \ref{app:DSwavebasis}). The solution then takes the form
\begin{equation}
    \Psi(z_{\ast})=r(z_{\ast})\Big[C_1W_{\text{in}}(z_{\ast})+C_2W_{\text{out}}(z_{\ast})\Big].
\end{equation}
Imposing $C_2=0$, corresponding to the absence of an outgoing wave from the throat, we obtain the solution describing the sub-throat modes
\begin{equation}
    R(r)=\left(\dfrac{r}{r_s}\right)^{ip}\Big[u^{(0)}_{-}(r)+2\chi\eta^{-\frac{1}{2}}u_{+}^{(0)}(r)\Big],
    \label{DSWHsubthroatsolution}
\end{equation}
where we have returned to the original function $R(r)$ and have also written the solution in terms of the even/odd basis, since the connection formulas \eqref{DSWHconection} are written for this basis. We can then expand the solution in the intermediate zone, where it can be separated into growing and decaying terms
\begin{equation}
    \begin{aligned}
        R^{\text{(R)}}(r)=\left(1-\dfrac{r_s}{r}\right)^{-ip}\left(C^{(0)+}_{-(\infty)}-2i\chi\eta^{-\frac{1}{2}}C^{(0)+}_{+(\infty)}\right)\left(\frac{r}{r_s}\right)^l\Bigg[
        &Z^{\text{source}}(r)+k_{\text{Love}}\left(\frac{r_s}{r}\right)^{2l+1}Z^{\text{response}}(r)\Bigg],
    \end{aligned}
    \label{DSWHsubthoatsolutioninfty}
\end{equation}
where $Z^{\text{source}}$ and $Z^{\text{response}}$ are given by \eqref{sourceresponsesplitdynDS}, and the Love numbers for the sub-throat modes are
\begin{equation}
    k_{\text{Love}}=\dfrac{C^{(0)-}_{-(\infty)}-2i\chi\eta^{-\frac{1}{2}}C^{(0)-}_{+(\infty)}}{C^{(0)+}_{-(\infty)}-2i\chi\eta^{-\frac{1}{2}}C^{(0)+}_{+(\infty)}}.
    \label{DSWHsubthroatalmostLove}
\end{equation}
Since propagating waves can exist in the throat only when $\chi=\sqrt{\omega^2r^2_s-\frac{\eta}{2}(1+2l(l+1))}$ is real, we expect the imaginary part of the Love numbers to vanish once $\chi$ vanishes or becomes imaginary. Indeed, all the connection coefficients (see appendix \ref{app:connection}) are real-valued functions, and therefore the Love numbers \eqref{DSWHsubthroatalmostLove} acquire an imaginary part only for real $\chi$, confirming this expectation.

Recall that in the previous calculations we analytically continued $l$ away from positive integer values. We must now return $l$ to positive integer values by making the shift $l\rightarrow l-\delta$, which allows us to isolate the resulting divergences. Using the explicit expressions for the connection coefficients, we find that the divergent part takes the form 
\begin{equation}
    k^{\text{(div.)}}_{\text{Love}}=-\eta\dfrac{(l!)^2l(l+1)}{4(2l)!(2l+1)!}\prod_{k=1}^{l}\left(k^2+4p^2\right)\dfrac{1}{\delta}.
\end{equation}
As in the case of the super-throat modes \eqref{DSWHdivfinite}, this divergence is canceled in the full solution \eqref{DSWHsubthoatsolutioninfty} by the corresponding divergence in the source function $Z^{\text{source}}(r)$ (see appendix \ref{app:zsource}). The complete solution is therefore regular. Thus, the Love numbers for the sub-throat modes exhibit the same RG flow as those for the super-throat modes, and therefore have the same classical $\beta$-function \eqref{DSWHbetafunction}. As a result, the Love numbers run with the scale and acquire an additional logarithmic contribution
\begin{equation}
    k^{(\text{ren.})}_{\text{Love}}=\eta\dfrac{(l!)^2l(l+1)}{2(2l)!(2l+1)!}\prod_{k=1}^{l}\left(k^2+4p^2\right)\ln\dfrac{\mu}{\mu_0}+k^{(\text{finite})}.
\end{equation}
We choose $\mu=\frac{1}{r}$ and $\mu_0=\frac{1}{r_s}$ and also retain the finite part. The sub-throat modes considered in this section, by definition, satisfy $\omega r_s\ll 1\ll \omega r_s\ln\frac{4}{\eta}$ and correspond to propagating modes well above the threshold $\left(\omega r_s\right)^2=\eta\left(l(l+1)+\frac{1}{2}\right)$. Therefore, after carrying out the calculation, we obtain the following expressions in $\eta$, $\frac{\eta}{\omega^2r_s^2}$ and $\frac{\eta}{\omega r_s}$ for the real and imaginary parts
\begin{equation}
    \begin{aligned}
        \Re k_{\text{Love}}
        &=-\eta\dfrac{(l!)^4l(l+1)}{2(2l)!(2l+1)!}\ln\dfrac{r}{r_s}+\eta\dfrac{(l!)^4}{(2l)!(2l+1)!}\Bigg[l(l+1)\sin\left(2\omega r_s\left(\ln\frac{\eta}{4}+2H_l\right)\right)+
        \\
        &+\dfrac{l(l+1)}{2}\left(2H_l-H_{2l+1}-H_{2l}\right)-\dfrac{1}{4\omega r_s}\left(1+l(l+1)\right)\sin\left(2\omega r_s\left(\ln\frac{\eta}{4}+2H_l\right)\right)\Bigg]
        \\
        \Im k_{\text{Love}}
        &=\dfrac{(l!)^4}{(2l)!(2l+1)!}\omega r_s\Bigg[1+\eta\Bigg(\dfrac{1}{2\omega^2r_s^2}\left(1+l(l+1)\right)\sin^2\left(\omega r_s\left(\ln\frac{\eta}{4}+2H_l\right)\right)+l^2-
        \\
        &-l(l+1)\left(\dfrac{\pi^2}{6}+H^{(2)}_l\right)-2l(l+1)\sin^2\left(\omega r_s\left(\ln\frac{\eta}{4}+2H_l\right)\right)-\dfrac{1+2l(l+1)}{4\omega^2r_s^2}\Bigg)\Bigg],
    \end{aligned}
    \label{DSWHsubthroatLove}
\end{equation}
which are valid, since
\begin{equation}
    \dfrac{\eta}{\left(\omega r_s\right)^2}\ll\eta\ln^2\dfrac{4}{\eta}\ll 1,\quad\text{and}\quad \dfrac{\eta}{\omega r_s}\ll\eta\ln\dfrac{4}{\eta}\ll 1.
\end{equation}
Recall that $\eta=\frac{\lambda^2}{1+\lambda^2}$ is the deformation parameter introduced in Eq.~\eqref{DSWHmetric}. It is important to note that, for the sub-throat modes, we cannot expand trigonometric functions in powers of $\omega r_s (\ln\frac{\eta}{4}+2H_l)$, since $\omega r_s\ll 1\ll \omega r_s\ln\frac{4}{\eta}$. We therefore keep all such functions in their exact form; however, we have expanded the other expressions in powers of $\omega r_s$. Although these trigonometric functions do not have a well-defined limit as $\eta\rightarrow 0$, they always appear multiplied by $\eta$. Since the sine and cosine functions are bounded, \eqref{DSWHsubthroatLove} has a well-defined limit as $\eta\rightarrow 0$.

\section{Static Love numbers and black-hole limit}\label{sec:staticandblack-hole}

To obtain the static Love numbers for both geometries under consideration, one should take the limit $\omega r_s\rightarrow 0$ in the expressions for the super-throat modes \eqref{SchWHLoveReIm} and \eqref{DSWHsuperthroatLovemodes}. Indeed, the static limit belongs to the super-throat case and can be viewed, roughly speaking, as a mode with an infinite wavelength, which is clearly much larger than the wormhole throat length. The Love numbers for the thin-shell (TS) wormhole and the DS wormhole then take the following forms, respectively
\begin{equation}
    \begin{aligned}
        k^{\text{TS}}_{\text{Love}}
        &=\dfrac{(l!)^4}{2(2l)!(2l+1)!}\dfrac{1}{\ln\epsilon+2H_l}\left[1+\epsilon\left((l(l+1)+1)\left(\ln\epsilon+2H_l\right)+\dfrac{1+2l(l+1)}{\ln\epsilon+2H_l}\right)\right]
        \\[5pt]
        k^{\text{DS}}_{\text{Love}}
        &=
        \dfrac{(l!)^4}{(2l)!(2l+1)!}\Bigg\{-\eta\dfrac{l(l+1)}{2}\ln\dfrac{r}{r_s}+\dfrac{1}{2\left(\ln\frac{\eta}{4}+2H_l\right)}\left(1-\dfrac{\eta}{2}\dfrac{1+(2l+1)\left(\ln\frac{\eta}{4}+2H_l\right)}{\ln\frac{\eta}{4}+2H_l}\right)+
        \\
        &+\eta\dfrac{l(l+1)}{4}\Bigg[\dfrac{1}{3}\left(\ln\frac{\eta}{4}+2H_l\right)^2+\left(\ln\frac{\eta}{4}+2H_l\right)\left(\ln\frac{\eta}{4}+6H_l-2H_{2l+1}-2H_{2l}\right)-2H^{(2)}_l-\dfrac{\pi^2}{3}-
        \\
        &-\dfrac{2H^{(3)}_l+6\zeta(3)}{3\left(\ln\frac{\eta}{4}+2H_l\right)}\Bigg]\Bigg\}.
    \end{aligned}
\end{equation}
An important difference between them should be emphasized: the static Love numbers of the DS wormhole exhibit running, just as their dynamical counterparts \eqref{DSWHsuperthroatLovemodes} do, and therefore their values depend on the observation scale. Apart from this difference, the static Love numbers show the same qualitative behavior in both geometries: their leading contribution scales as the inverse logarithm of the deformation parameter, in full agreement with previous studies of black-hole mimickers.

Another important limit is the black-hole limit. When approaching a black hole, $\epsilon\rightarrow 0$ for the thin-shell wormhole and $\eta\rightarrow 0$ for the DS wormhole. In this limit, the wormhole throat becomes infinitely long, while the range of super-throat modes shrinks to zero, so that these modes gradually become irrelevant. At the same time, the sub-throat modes become the relevant ones. Therefore, to analyze the black-hole limit, one should consider the sub-throat modes, which, as the deformation parameters approach zero, recover the standard dynamical Love numbers of a Schwarzschild black hole at first order in $\omega r_s$ \cite{Combaluzier--Szteinsznaider:2025eoc, Kobayashi:2025vgl}
\begin{equation}
    \Re k_{\text{Love}}^{\text{(BH)}}=0,\quad \Im k_{\text{Love}}^{\text{(BH)}}=\omega r_s\dfrac{(l!)^4}{(2l)!(2l+1)!},
\end{equation}
so there is a smooth analytic transition from the wormhole Love numbers to the black hole Love numbers.

\section{Discussion and Conclusions}\label{conclusions}

In this work, we studied the scalar tidal Love numbers of two spherically symmetric wormhole geometries: a thin-shell wormhole, constructed by gluing together two copies of the Schwarzschild geometry, and the Damour-Solodukhin (DS) wormhole, which differs from the Schwarzschild metric by a small deformation parameter that removes the horizon. From a topological perspective, the two wormholes are similar: both connect two asymptotically flat universes and mimic a black hole in each universe. Consequently, their effective potentials exhibit a characteristic double-peak structure, with the two peaks separated by a distance that depends logarithmically on the parameter controlling the deviation from the black-hole geometry.
The two geometries, however, differ in their smoothness properties at the throat. By construction, the thin-shell wormhole is obtained by continuously gluing two Schwarzschild geometries at the throat, but the resulting metric is not smooth there. Away from the throat, the geometry coincides exactly with the Schwarzschild metric. In contrast, the DS wormhole is smooth across the throat, while its metric differs from the Schwarzschild geometry by a small deformation proportional to the parameter  $\eta$ outside the throat.

In studying dynamical Love numbers, we assume a monochromatic external tidal field characterized by a frequency $\omega$. For a black hole, the corresponding wavelength can be compared with the characteristic size of the black hole, naturally separating the problem into low- and high-frequency regimes. For a wormhole, the presence of a long throat introduces an additional length scale, namely the throat length. This allows us to distinguish three classes of modes according to their wavelengths (or, equivalently, their frequencies): short-wavelength (high-frequency), sub-throat, and super-throat modes, with the latter two belonging to the low-frequency regime.
Dividing the low-frequency regime into two subclasses makes the interpretation of the Love numbers somewhat more subtle. Our calculations are performed for monochromatic modes. Nevertheless, this separation provides a useful connection to the well-known properties of wormhole ringdown \cite{Cardoso:2016oxy, Hui:2019aox}. In particular, the early-time ringdown signal of a wormhole can closely reproduce that of a black hole, with deviations becoming apparent only through the subsequent echoes at observation times larger than approximately twice the throat length. At the same time, the true quasinormal modes (QNMs) of the wormhole \cite{Solodukhin:2025opw, Bueno:2017hyj}, which govern its complete ringdown response, do not contain the black-hole QNMs and do not approach them analytically in the black-hole limit.

Future gravitational-wave observations are expected to provide information about the Love numbers of compact objects. If the observed object is a wormhole with a very long throat—or, equivalently, with a very small deformation away from a black hole, as may be expected if the deformation is of quantum origin—the observer will not immediately receive a signal from the second universe. During this time interval, the Love numbers inferred from the gravitational-wave signal will therefore correspond to the sub-throat modes.

In this work, we computed the scalar Love numbers for sub-throat modes for both wormhole geometries. To impose the appropriate boundary conditions, we constructed bases of ingoing and outgoing waves at the throat in each case. Owing to the similar topology of the two geometries, the resulting Love numbers share several universal properties. At leading order, the real part is linear in the deformation parameter. The imaginary part begins at linear order in $\omega r_s$, as in the black-hole case, and contains both a finite contribution at zeroth order in the deformation parameter and a first-order correction. Thus, in the black-hole limit, the sub-throat scalar Love numbers continuously approach their corresponding Schwarzschild black-hole values. Another important feature of the sub-throat modes is the presence of a threshold for wave propagation inside the wormhole throat. In both geometries, the corresponding threshold is encoded in the unexpanded expressions for the Love numbers \eqref{SchWHsubthroatalmostLove} and \eqref{DSWHsubthroatalmostLove}, but does not appear explicitly in the approximate results \eqref{SchWHsubthroatLovenumbers} and \eqref{DSWHsubthroatLove}. The reason is that the approximate expressions are derived for modes lying parametrically well above the threshold. Indeed, the sub-throat modes are defined by $\omega r_s\ll 1\ll \omega r_s\ln\frac{1}{\epsilon}$ for the thin-shell wormhole and by $\omega r_s\ll 1\ll \omega r_s\ln\frac{4}{\eta}$ for the DS wormhole. In the small-deformation limit, these conditions ensure that the relevant modes lie well within the propagating regime. The threshold behavior is therefore not visible in the expanded expressions, although it remains present in the underlying unexpanded expressions for the Love numbers.

There are, however, important differences arising from the distinct smoothness properties of the two geometries. Most notably, the scalar Love numbers of the DS wormhole exhibit logarithmic running. This behavior signals the presence of a classical renormalization-group (RG) flow of the Wilson coefficients encoding finite-size effects beyond the point-particle approximation. 

We also computed the Love numbers describing the full wormhole response, which is sensitive to both asymptotically flat universes and corresponds to the super-throat modes. These Love numbers exhibit several universal features. At leading order, the real part, corresponding to the static Love numbers, displays the inverse-logarithmic behavior previously found in the literature \cite{Cardoso:2017cfl},
\begin{equation}
\Re k_{\text{Love}}^{(\text{thin-shell})}\sim \dfrac{1}{\ln\epsilon+2H_l},\quad
\Re k_{\text{Love}}^{(\text{DS})}\sim \dfrac{1}{\ln\frac{\eta}{4}+2H_l}.
\end{equation}
We also calculated the subleading corrections for both geometries. In the black-hole limit, the real part vanishes, consistently with the vanishing static Love numbers of a Schwarzschild black hole \cite{Fang:2005qq, Binnington:2009bb, Damour:2009vw, Kol:2011vg, Hui:2020xxx}.

Unlike the sub-throat modes, for which the dissipative part of the Love numbers appears at first order in $\omega r_s$ \eqref{SchWHsubthroatLovenumbers} and \eqref{DSWHsubthroatLove}, consistent with the analogous behavior of black-hole Love numbers \cite{Combaluzier--Szteinsznaider:2025eoc, Kobayashi:2025vgl}, the dissipative part of the super-throat modes appears only at order $(\omega r_s)^{2l+1}$, with the leading contribution proportional to the inverse square of the logarithm of the deformation parameter,
\begin{equation}
\Im k_{\text{Love}}^{(\text{thin-shell})}\sim
\dfrac{(\omega r_s)^{2l+1}}{\left(\ln\epsilon+2H_l\right)^2},\quad
\Im k_{\text{Love}}^{(\text{DS})}\sim
\dfrac{(\omega r_s)^{2l+1}}{\left(\ln\frac{\eta}{4}+2H_l\right)^2}.
\end{equation}
We also obtained the corresponding subleading corrections. This difference allows the two types of modes to be distinguished. Consequently, in the black-hole limit, the imaginary part of the super-throat Love numbers vanishes and therefore does not reproduce the Schwarzschild result.

This difference can be understood from the global structure probed by the super-throat modes. These modes interact with both asymptotically flat universes, each of which contains its own centrifugal potential barrier. In contrast, the Schwarzschild black-hole potential contains only a single such barrier. The presence of the second barrier therefore introduces an additional low-frequency suppression of the dissipative response.

As in the sub-throat case, the super-throat scalar Love numbers of the DS wormhole contain a distance-dependent logarithmic contribution and consequently exhibit logarithmic running. Thus, despite describing physically distinct tidal responses and being derived using different boundary conditions, both classes of Love numbers exhibit a classical RG flow of the Wilson coefficients encoding finite-size effects.

The full wormhole scalar Love numbers, corresponding to the super-throat modes, therefore display behavior analogous to that of the wormhole QNMs. They do not possess a smooth analytic limit that reproduces the corresponding black-hole quantities. In particular, the Schwarzschild Love numbers are not recovered simply by taking the deformation parameter to zero. This can be understood physically from the behavior of the wormhole throat: as the black-hole limit is approached, the throat becomes increasingly long. Consequently, the range of super-throat modes shrinks, and these modes gradually become irrelevant, disappearing entirely in the limiting case. At the same time, the sub-throat modes become increasingly important and continuously recover the Love numbers of the Schwarzschild black hole. We also note that the potential instability associated with the radial perturbations of thin shells \cite{Poisson:1995sv} and with particle collisions near the throat \cite{Tsukamoto:2019ihj} is an interesting issue that deserve further investigation.

Wormholes therefore exhibit a nontrivial tidal response associated with the presence of a long throat, closely paralleling the phenomena already identified in their ringdown signals. Although our analysis is restricted to scalar perturbations, our results demonstrate that dynamical Love numbers provide a valuable probe of the structure of compact objects and can reveal features that distinguish wormholes from black holes. In particular, the existence of distinct sub-throat and super-throat responses highlights the role of the global geometry in determining the tidal response and suggests that Love numbers may provide a promising signature of deviations from the standard black-hole paradigm.

\section*{Acknowledgments} 

The authors would like to thank E. Babichev, O. Lisovyy and F. Vernizzi for useful discussions.

\vspace{0.5cm}
\section*{Appendix}
\appendix
\section{Heun equation}
\label{app:HeunEq}

The Heun equation has the following form
\begin{equation}
     \dfrac{\text{d}^2}{\text{d}z^2}u(z) + \left(\dfrac{\gamma}{z}+\dfrac{\delta}{z-1}+\dfrac{\varepsilon}{z-\eta}\right)\dfrac{\text{d}}{\text{d}z}u(z)+\dfrac{\alpha\beta z-q}{z(z-1)(z-\eta)}u(z)=0,
     \label{Heuneq}
\end{equation}
with $1+\alpha+\beta=\gamma+\delta+\varepsilon$. All its solutions can be written in terms of a function that admits a power-series expansion around $0$, which converges in a disk $|z|<\text{min}(1,|\eta|)$
\begin{equation}
    \text{HeunG}(\eta,q|\alpha,\beta,\gamma,\delta|z) = \sum_{k=0}^{\infty}c_kz^k.
\end{equation}
The coefficients satisfy a recurrence relation
\begin{gather}
    R_{k}c_{k+1}-(q+Q_{k})c_{k}+P_kc_{k-1}=0,\quad c_0=1,\quad c_{-1}=0\nonumber\\
    R_k=\eta(k+1)(k+\gamma)\nonumber\\
    Q_k=k\left((k-1+\gamma)(1+\eta)+\eta\delta+\epsilon\right)\\
    P_k=(k-1+\alpha)(k-1+\beta).\nonumber
\end{gather}
A problem may arise if $\gamma$ is a negative integer. The coefficients $c_k$ can be written as 
\begin{equation}
    c_k=\dfrac{\Delta_k}{\prod_{j=0}^{k-1}R_j} = \dfrac{\Gamma(\gamma)\Delta_k}{\eta^k\Gamma(\gamma+k)k!},
    \label{coefheun4}
\end{equation}
where $\Delta_k$ satisfies a new recurrence relation
\begin{equation}
    \Delta_k=\left(q+Q_{k-1}\right)\Delta_{k-1}-P_{k-1}R_{k-2}\Delta_{k-2},\quad \Delta_0=1,\quad\Delta_{-1}=0.
    \label{recrelationHeun4}
\end{equation}

By introducing the function $\psi(z)=z^{\gamma/2}(z-1)^{\delta/2}(z-\eta)^{\varepsilon/2}u(z)$, the Heun equation can be brought into its normal form
\begin{equation}
    \dfrac{\text{d}^2}{\text{d}z^2}\psi(z)+\left[\dfrac{\frac{1}{4}-a_0^2}{z^2}+\dfrac{\frac{1}{4}-a_1^2}{(z-1)^2}+\dfrac{\frac{1}{4}-a_\eta^2}{(z-\eta)^2}+\dfrac{-\frac{1}{2}+a_0^2+a_1^2+a_\eta^2-a_{\infty}^2}{z(z-1)}+\dfrac{(\eta-1)u^{(0)}}{z(z-1)(z-\eta)}\right]\psi(z)=0,
    \label{HeunnormalForm}
\end{equation}
where the coefficients are related to those of the original equation \eqref{Heuneq} by
\begin{equation}
    \begin{aligned}
        a_0=\dfrac{1-\gamma}{2},\quad a_1=\dfrac{1-\delta}{2},\quad a_{\eta}=\dfrac{1-\varepsilon}{2},\quad a_{\infty}=\dfrac{\alpha-\beta}{2},\quad u^{(0)}=\dfrac{-2q+2\eta\alpha\beta+\gamma\varepsilon-\eta(\gamma+\delta)\varepsilon}{2(\eta-1)}
    \end{aligned}
    \label{HeunNormalFormParameters}
\end{equation}

\section{Connection coefficients}\label{app:connection}

We now use a recent result \cite{Bonelli:2022ten} to write down the connection coefficients relevant to the problem considered in section \ref{sec:DSWH}. For the connection between the solutions of Eq.~\eqref{Heuneq} around $z=0$ and $z=\infty$, these coefficients take the form 
\begin{equation}
    C^{(0)\theta_1}_{\theta(\infty)} = \eta^{\frac{1}{2}-a_\eta+\theta a_0}e^{\frac{\theta}{2}\partial_{a_0}F-\frac{\theta_1}{2}\partial_{a_{\infty}}F}\left(\sum_{\sigma=\pm}M_{\theta\sigma}(a_0,a,a_\eta)M_{(-\sigma)\theta_1}(a,a_{\infty},a_1)\eta^{\sigma a}e^{-\frac{\sigma}{2}\partial_aF}\right)
    \label{connectionformula}
\end{equation}
where 
\begin{equation}
    \begin{aligned}
        &F(\eta)=\dfrac{\left(\frac{1}{4}-a^2-a_1^2+a_{\infty}^2\right)\left(\frac{1}{4}-a^2-a_{\eta}^2+a_{0}^2\right)}{\frac{1}{2}-2a^2}\eta + \mathcal{O}(\eta^2),
        \\
        &M_{\theta_1\theta_2}(a,b,c)=\frac{\Gamma(1+2\theta_1a)\Gamma(-2\theta_2b)}{\Gamma(\frac{1}{2}+\theta_1a-\theta_2b+c)\Gamma(\frac{1}{2}+\theta_1a-\theta_2b-c)},
    \end{aligned}
\end{equation}
$a_0$, $a_1$, $a_\eta$ and $a_{\infty}$ are parameters of the normal form \eqref{HeunnormalForm}, while $a$ is an auxiliary parameter that can be determined perturbatively from the equation
\begin{equation}
    u^{(0)}=-\dfrac{1}{4}-a^2+a_{\eta}^2+a_0^2+\eta\partial_{\eta}F(\eta).
    \label{ConnectionFormulaAuxiliaryEquation}
\end{equation}

For the DS wormhole considered in this work (see section \ref{sec:DSWH}), these parameters are given by
\begin{equation}
    a_0=\dfrac{1}{4},\quad a_1=-ip,\quad a_{\eta}=\dfrac{1}{4},\quad a_{\infty}=\dfrac{-2l-1}{2},\quad u^{(0)}=\dfrac{1-3\eta-8p^2}{8(\eta-1)}.
\end{equation}
Solving the equation \eqref{ConnectionFormulaAuxiliaryEquation} up to first order in $\eta$, we obtain
\begin{equation}
    a^2=-p^2+\dfrac{\eta}{2}l(l+1)+\mathcal{O}(\eta^2)
\end{equation}
and the connection coefficients \eqref{connectionformula} become
\begin{equation}
    \begin{aligned}
        &C^{(0)-}_{-(\infty)}=e^{-\frac{1}{2}\partial_{a_0}F+\frac{1}{2}\partial_{a_{\infty}}F}\dfrac{\Gamma(-2l-1)}{2}\sum_{\sigma=\pm}\left(\frac{\eta}{4}\right)^{\sigma a}e^{-\frac{\sigma}{2}\partial_{a}F(\eta)}\dfrac{\Gamma(1-2\sigma a)}{\Gamma(-l-ip-\sigma a)\Gamma(-l+ip-\sigma a)}
        \\
        &C^{(0)+}_{-(\infty)}=e^{-\frac{1}{2}\partial_{a_0}F-\frac{1}{2}\partial_{a_{\infty}}F}\dfrac{\Gamma(2l+1)}{2}\sum_{\sigma=\pm}\left(\frac{\eta}{4}\right)^{\sigma a}e^{-\frac{\sigma}{2}\partial_{a}F(\eta)}\dfrac{\Gamma(1-2\sigma a)}{\Gamma(l+1-ip-\sigma a)\Gamma(l+1+ip-\sigma a)}
        \\
        &C^{(0)-}_{+(\infty)}=-\eta^{\frac{1}{2}}e^{\frac{1}{2}\partial_{a_0}F+\frac{1}{2}\partial_{a_{\infty}}F}\dfrac{\Gamma(-2l-1)}{4}\sum_{\sigma=\pm}\dfrac{\sigma}{a}\left(\frac{\eta}{4}\right)^{\sigma a}e^{-\frac{\sigma}{2}\partial_{a}F(\eta)}\dfrac{\Gamma(1-2\sigma a)}{\Gamma(-l-ip-\sigma a)\Gamma(-l+ip-\sigma a)}
        \\
        &C^{(0)+}_{+(\infty)}=-\eta^{\frac{1}{2}}e^{+\frac{1}{2}\partial_{a_0}F-\frac{1}{2}\partial_{a_{\infty}}F}\dfrac{\Gamma(2l+1)}{4}\sum_{\sigma=\pm}\dfrac{\sigma}{a}\left(\frac{\eta}{4}\right)^{\sigma a}e^{-\frac{\sigma}{2}\partial_{a}F(\eta)}\dfrac{\Gamma(1-2\sigma a)}{\Gamma(l+1-ip-\sigma a)\Gamma(l+1+ip-\sigma a)}
    \end{aligned}
    \label{connectionDSWH1}
\end{equation}

We must now expand these expressions to first order in $\eta$. This yields the following rather lengthy formulas
\begin{equation}
    \begin{aligned}
        C^{(0)-}_{-(\infty)}
        & =e^{-\frac{1}{2}\partial_{a_0}F+\frac{1}{2}\partial_{a_{\infty}}F}\dfrac{\Gamma(-2l-1)}{2\Gamma(-l)}\left[\sum_{\sigma=\pm}\left(\dfrac{\eta}{4}\right)^{\sigma ip}\dfrac{\Gamma(1-2\sigma ip)}{\Gamma(-l-2\sigma ip)}+\dfrac{\eta}{2}ip\sum_{\sigma=\pm}\sigma\left(\dfrac{\eta}{4}\right)^{\sigma ip}\dfrac{\Gamma(1-2\sigma ip)}{\Gamma(-l-2\sigma ip)}+\right.
        \\
        &\left.+\eta\dfrac{l(l+1)}{4ip}\sum_{\sigma=\pm}\sigma\left(\dfrac{\eta}{4}\right)^{\sigma ip}\dfrac{\Gamma(1-2\sigma ip)}{\Gamma(-l-2\sigma ip)}\left(\ln\dfrac{\eta}{4}+\psi(-l)-2\psi(1-2\sigma ip)+\psi(-l-2\sigma ip)\right)\right]
        \\
        C^{(0)+}_{-(\infty)}
        & =e^{-\frac{1}{2}\partial_{a_0}F-\frac{1}{2}\partial_{a_{\infty}}F}\dfrac{\Gamma(2l+1)}{2\Gamma(l+1)}\left[\sum_{\sigma=\pm}\left(\dfrac{\eta}{4}\right)^{\sigma ip}\dfrac{\Gamma(1-2\sigma ip)}{\Gamma(l+1-2\sigma ip)}+\dfrac{\eta}{2}ip\sum_{\sigma=\pm}\sigma\left(\dfrac{\eta}{4}\right)^{\sigma ip}\dfrac{\Gamma(1-2\sigma ip)}{\Gamma(l+1-2\sigma ip)}+\right.
        \\
        &\left.+\eta\dfrac{l(l+1)}{4ip}\sum_{\sigma=\pm}\sigma\left(\dfrac{\eta}{4}\right)^{\sigma ip}\dfrac{\Gamma(1-2\sigma ip)}{\Gamma(l+1-2\sigma ip)}\left(\ln\dfrac{\eta}{4}+\psi(l+1)-2\psi(1-2\sigma ip)+\psi(l+1-2\sigma ip)\right)\right]
        \\
        C^{(0)-}_{+(\infty)}
        & =-\dfrac{1}{4}\eta^{\frac{1}{2}}e^{\frac{1}{2}\partial_{a_0}F+\frac{1}{2}\partial_{a_{\infty}}F}\dfrac{\Gamma(-2l-1)}{\Gamma(-l)}\left[\dfrac{1}{ip}\sum_{\sigma=\pm}\sigma\left(\dfrac{\eta}{4}\right)^{\sigma ip}\dfrac{\Gamma(1-2\sigma ip)}{\Gamma(-l-2\sigma ip)}+\right.  
        \\
        &+\dfrac{\eta}{2}\sum_{\sigma=\pm}\left(\dfrac{\eta}{4}\right)^{\sigma ip}\dfrac{\Gamma(1-2\sigma ip)}{\Gamma(-l-2\sigma ip)}+\eta\dfrac{l(l+1)}{4ip^3}\sum_{\sigma=\pm}\sigma\left(\dfrac{\eta}{4}\right)^{\sigma ip}\dfrac{\Gamma(1-2\sigma ip)}{\Gamma(-l-2\sigma ip)}-
        \\
        &\left.-\eta\dfrac{l(l+1)}{4p^2}\sum_{\sigma=\pm}\left(\dfrac{\eta}{4}\right)^{\sigma ip}\dfrac{\Gamma(1-2\sigma ip)}{\Gamma(-l-2\sigma ip)}\left(\ln\dfrac{\eta}{4}+\psi(-l)-2\psi(1-2\sigma ip)+\psi(-l-2\sigma ip)\right)\right]
        \\
        C^{(0)+}_{+(\infty)}
        & =-\dfrac{1}{4}\eta^{\frac{1}{2}}e^{\frac{1}{2}\partial_{a_0}F-\frac{1}{2}\partial_{a_{\infty}}F}\dfrac{\Gamma(2l+1)}{\Gamma(l+1)}\left[\dfrac{1}{ip}\sum_{\sigma=\pm}\sigma\left(\dfrac{\eta}{4}\right)^{\sigma ip}\dfrac{\Gamma(1-2\sigma ip)}{\Gamma(l+1-2\sigma ip)}+\right. 
        \\
        &+\dfrac{\eta}{2}\sum_{\sigma=\pm}\left(\dfrac{\eta}{4}\right)^{\sigma ip}\dfrac{\Gamma(1-2\sigma ip)}{\Gamma(l+1-2\sigma ip)}+\eta\dfrac{l(l+1)}{4ip^3}\sum_{\sigma=\pm}\sigma\left(\dfrac{\eta}{4}\right)^{\sigma ip}\dfrac{\Gamma(1-2\sigma ip)}{\Gamma(l+1-2\sigma ip)}-
        \\
        &\left.-\eta\dfrac{l(l+1)}{4p^2}\sum_{\sigma=\pm}\left(\dfrac{\eta}{4}\right)^{\sigma ip}\dfrac{\Gamma(1-2\sigma ip)}{\Gamma(l+1-2\sigma ip)}\left(\ln\dfrac{\eta}{4}+\psi(l+1)-2\psi(1-2\sigma ip)+\psi(l+1-2\sigma ip)\right)\right]
    \end{aligned}
\end{equation}

Recall that throughout the calculation we have treated $l$ as a non-integer number. We now restore positive integer values of $l$ by making the shift $l\rightarrow l-\delta$, which allows us to isolate possible divergences in $\delta$. Two of the connection coefficients contain such divergences and take the following form
\begin{equation}
    \begin{aligned}
        &C^{(0)-}_{-(\infty)}=\dfrac{1}{\delta}\left[C^{(0)-}_{-(\infty)}\right]^{\text{sing.}}+\left[C^{(0)-}_{-(\infty)}\right]^{\text{reg.}}
        \\
        &\left[C^{(0)-}_{-(\infty)}\right]^{\text{sing.}}=-\eta\dfrac{l!l(l+1)}{4(2l+1)!}\Theta^{-1}_l\cos(px)
        \\
        &\left[C^{(0)-}_{-(\infty)}\right]^{\text{reg.}}=\dfrac{l!\Theta^{-1}_le^{-\frac{1}{2}\partial_{a_0}F+\frac{1}{2}\partial_{a_{\infty}}F}}{(2l+1)!}\Bigg[-p\sin(px)-\dfrac{\eta}{2}p^2\cos(px)+\eta\dfrac{l(l+1)}{4}\left(U_l\cos(px)+V_l\sin(px)\right)\Bigg]
    \end{aligned}
    \label{concoef1}
\end{equation}
and
\begin{equation}
    \begin{aligned}
        &C^{(0)-}_{+(\infty)}=\dfrac{1}{\delta}\left[C^{(0)-}_{+(\infty)}\right]^{\text{sing.}}+\left[C^{(0)-}_{+(\infty)}\right]^{\text{reg.}}
        \\
        &\left[C^{(0)-}_{+(\infty)}\right]^{\text{sing.}}=\eta^{\frac{3}{2}}\dfrac{l!l(l+1)}{8(2l+1)!}\Theta^{-1}_l\dfrac{1}{p}\sin(px)
        \\
        &\left[C^{(0)-}_{+(\infty)}\right]^{\text{reg.}}=-\eta^{\frac{1}{2}}\dfrac{l!\Theta^{-1}_le^{\frac{1}{2}\partial_{a_0}F+\frac{1}{2}\partial_{a_{\infty}}F}}{2(2l+1)!}\Bigg[\cos(px)-\dfrac{\eta}{2}p\sin(px)+\eta\dfrac{l(l+1)}{4p}\left(U_l\sin(px)-V_l\cos(px)\right)+
        \\
        &+\eta\dfrac{l(l+1)}{4p^2}\cos(px)\Bigg],
    \end{aligned}
    \label{concoef2}
\end{equation}
where we have introduced $\Theta_l=\prod_{k=1}^{l}\left(k^2+4p^2\right)^{-\frac{1}{2}}$, $x=\ln\frac{\eta}{4}+\frac{1}{p}\sum_{k=1}^{l}\arctan\frac{2p}{k}$ and
\begin{equation}
    U_l=\ln\dfrac{\eta}{4}+2\psi(l+1)-2\psi(2l+2)+2\sum_{k=1}^l\dfrac{k}{k^2+4p^2},\quad V_l=\dfrac{1}{p}+4p\sum_{k=1}^l\dfrac{1}{k^2+4p^2}.
\end{equation}

The remaining two coefficients are finite. We therefore expand them to first order in $\delta$. Since only the zeroth order contribution in $\eta$ of $\left[C^{(0)+}_{\pm(\infty)}\right]^{(1)}$ will be needed, we display only those terms
\begin{equation}
    \begin{aligned}
        &C^{(0)+}_{-(\infty)}=\left[C^{(0)+}_{-(\infty)}\right]^{(0)}+\delta\left[C^{(0)+}_{-(\infty)}\right]^{(1)}
        \\
        &\left[C^{(0)+}_{-(\infty)}\right]^{(0)}=e^{-\frac{1}{2}\partial_{a_0}F-\frac{1}{2}\partial_{a_{\infty}}F}\dfrac{(2l)!}{l!}\Theta_l\Bigg[\cos(px)-\dfrac{\eta}{2}p\sin(px)+\eta\dfrac{l(l+1)}{4p}\Big(A_l\sin(px)+B_l\cos(px)\Big)\Bigg]
        \\
        &\left[C^{(0)+}_{-(\infty)}\right]^{(1)}=\dfrac{(2l)!}{l!}\Theta_l\Big[\tilde{A}_l\cos(px)-\tilde{B}_l\sin(px)\Big]+\mathcal{O}(\eta)
    \end{aligned}
    \label{concoef3}
\end{equation}
and 
\begin{equation}
    \begin{aligned}
        &C^{(0)+}_{+(\infty)}=\left[C^{(0)+}_{+(\infty)}\right]^{(0)}+\delta\left[C^{(0)+}_{+(\infty)}\right]^{(1)}
        \\
        &\left[C^{(0)+}_{+(\infty)}\right]^{(0)}=-\dfrac{\eta^{\frac{1}{2}}}{2}e^{\frac{1}{2}\partial_{a_0}F-\frac{1}{2}\partial_{a_{\infty}}F}\dfrac{(2l)!}{l!}\Theta_l\Big[\dfrac{1}{p}\sin(px)+\dfrac{\eta}{2}\cos(px)-\eta\dfrac{l(l+1)}{4p^2}\left(A_l\cos(px)-B_l\sin(px)\right)+
        \\
        &+\eta\dfrac{l(l+1)}{4p^3}\sin(px)\Bigg]
        \\
        &\left[C^{(0)+}_{+(\infty)}\right]^{(1)}=-\dfrac{\eta^{\frac{1}{2}}}{2p}\Theta_l\dfrac{(2l)!}{l!}\Big[\tilde{A}_l\sin(px)+\tilde{B}_l\cos(px)\Big]+\mathcal{O}(\eta^{\frac{3}{2}}),
    \end{aligned}
    \label{concoef4}
\end{equation}
where we have introduced the following functions
\begin{equation}
    \begin{aligned}
        &\tilde{A}_l=\Re\psi(1+2ip)+\psi(l+1)-2\psi(2l+1)+\sum_{k=1}^l\dfrac{k}{k^2+4p^2},\quad \tilde{B}_l=-\Bigg[\dfrac{\pi}{2}\coth(2\pi p)-\dfrac{1}{4p}\Bigg]+2p\sum_{k=1}^l\dfrac{1}{k^2+4p^2},
        \\
        &A_l=\ln\frac{\eta}{4}+\psi(l+1)-\Re\psi(1+2ip)+\sum_{k=1}^l\dfrac{k}{k^2+4p^2},\quad B_l=\Bigg[\dfrac{\pi}{2}\coth(2\pi p)-\dfrac{1}{4p}\Bigg]+2p\sum_{k=1}^l\dfrac{1}{k^2+4p^2}
    \end{aligned}
\end{equation}

It is worth noting that all connection coefficients are even functions of $p$ and hence of $\omega r_s$.

\section{Regularization of the $Z^{\text{source}}(r)$ function}\label{app:zsource}
As we noted in section \ref{sec:scalarfieldequationDSWH}, $Z^{\text{source}}(r)$ in Eq.~\eqref{sourceresponsesplitdynDS} is singular when $l$ is a positive integer, since $\gamma=-2l$; see Eq.~\ref{coefheun4}.
\begin{equation}
    \begin{aligned}
        &Z^{\text{source}}(r) = \left(1-\dfrac{r_s}{r}\right)^{l}\text{HeunG}\left(\eta,l^2-\dfrac{l}{2}(\eta+4ip)\left|-l+ip,-l+\dfrac{1}{2}+ip,-2l,\dfrac{1}{2}\right|\dfrac{\eta r_s}{r_s-r}\right) = 
        \\
        &=\left(1-\dfrac{r_s}{r}\right)^{l}\sum_{k=0}^{\infty}\dfrac{\Delta_k\Gamma(-2l)}{k!\Gamma(-2l+k)}\left(\dfrac{r_s}{r_s-r}\right)^k = \left(1-\dfrac{r_s}{r}\right)^l\sum_{k=0}^{2l}\dfrac{\Delta_k\Gamma(-2l)}{k!\Gamma(-2l+k)}\left(\dfrac{r_s}{r_s-r}\right)^k +
        \\
        &+\left(1-\dfrac{r_s}{r}\right)^{l}\sum_{k=2l+1}^{\infty}\dfrac{\Delta_k\Gamma(-2l)}{k!\Gamma(-2l+k)}\left(\dfrac{r_s}{r_s-r}\right)^k,
    \end{aligned}
    \label{almostrenormsource}
\end{equation}
where $\Delta_k$ satisfies a recurrence relation \eqref{recrelationHeun4}
\begin{equation}
    \Delta_k=\left(q+Q_{k-1}\right)\Delta_{k-1}-P_{k-1}R_{k-2}\Delta_{k-2},\quad \Delta_0=1,\quad\Delta_{-1}=0.
    \label{recnonrot}
\end{equation}
We split the series \eqref{almostrenormsource} into two parts. The first part is a finite sum and remains regular when $l$ is a positive integer, since $k\leq 2l$. The second part is proportional to the divergent contribution arising from the factor $\Gamma(-2l)$ in the numerator.

For our set of parameters, the coefficients take the form:
\begin{gather}
    R_{k}=\eta(k+1)(k-2l)\nonumber\\
    q+Q_k=(k-l)(k-l+2ip)+\eta\left[-\dfrac{l}{2}+k\left(k-2l-\dfrac{1}{2}\right)\right]\\
    P_{k}=(k-l-1+ip)\left(k-l-\dfrac{1}{2}+ip\right)\nonumber
\end{gather}

We solve for $\Delta_k$ up to first order in $\eta$. For this purpose, we introduce $\theta_k=\Delta_k/\Delta_{k-1}$
\begin{equation}
    \theta_k=q+Q_{k-1}-\dfrac{P_{k-1}R_{k-2}}{\theta_{k-1}}
\end{equation}
and expand it to first order in $\eta$, which gives the following expression
\begin{equation}
    \begin{aligned}
         \theta_k
         &=(k-l-1)(k-l-1+2ip)+
         \\
         &+\eta\left[-\dfrac{l}{2}+(k-1)\left(k-2l-\dfrac{3}{2}\right)-\dfrac{(k-1)(k-2-2l)(k-2-l+ip)\left(k-l-\frac{3}{2}+ip\right)}{(k-l-2)(k-l-2+2ip)}\right]+\mathcal{O}(\eta^2)
    \end{aligned}
\end{equation}
Then $\Delta_k$ can be written in terms of $\theta_k$ as follows:
\begin{equation}
    \begin{aligned}
        \Delta_k=\theta_k\theta_{k-1}\dots\theta_1\Delta_0 &=\prod_{j=1}^{k}\left[(j-l-1)(j-l-1+2ip)\right]\cdot
        \\
        &\cdot\left(1+\eta\left[\sum_{j=0}^{k-1}\dfrac{A_1}{j-l-1}+\sum_{j=0}^{k-1}\dfrac{A_2}{j-l}+\sum_{j=0}^{k-1}\dfrac{B_1}{j-l-1+2ip}+\sum_{j=0}^{k-1}\dfrac{B_2}{j-l+2ip}\right]\right)
    \end{aligned}
    \label{lndeltaDynDS}
\end{equation}
where
\begin{equation}
    \begin{aligned}
        &A_1=\dfrac{l(l+1)}{4},\quad A_2=-\dfrac{l(l+1)(1+ip)}{4ip}
        \\
        &B_1=\dfrac{l(l+1)+2ip(1-2ip)}{4},\quad B_2=\dfrac{l(l+1)(1-ip)+2p^2(1-2ip)}{4ip}
    \end{aligned}
\end{equation}
We also use the following useful identities involving the digamma and polygamma functions:
\begin{gather}
    \sum_{i=0}^{m}\dfrac{1}{i+z}=\psi(z+m+1)-\psi(z),\quad
    \sum_{i=0}^{m}\dfrac{1}{(i+z)^2}=\psi^{(1)}(z)-\psi^{(1)}(z+m+1)
\end{gather}
In our case, this allows us to rewrite the result of \eqref{lndeltaDynDS} as 
\begin{equation}
    \begin{aligned}
        \Delta_k
        &=\dfrac{\Gamma(-l+k)\Gamma(-l+k+2ip)}{\Gamma(-l)\Gamma(-l+2ip)}\left(1+\eta\left[A_1\left(\psi(-l-1+k)-\psi(-l-1)\right)+A_2\left(\psi(-l+k)-\psi(-l)\right)+\right.\right.
        \\
        &\left.\left.+B_1\left(\psi(-l-1+2ip+k)-\psi(-l-1+2ip)\right)+B_2\left(\psi(-l+2ip+k)-\psi(-l+2ip)\right)\right]\right)
    \end{aligned}
    \label{deltaDynDS}
\end{equation}

We now rearrange the second part in \eqref{almostrenormsource} into the following form
\begin{equation}
    -\dfrac{\Delta_{2l+1}\Gamma(-2l)}{\Gamma(2l+2)}\left(\dfrac{r_s}{r}\right)^{2l+1}\left(1-\dfrac{r_s}{r}\right)^{-l-1}\sum_{k=0}^{\infty}\dfrac{\Delta_{2l+1+k}/\Delta_{2l+1}\Gamma(2l+2)}{k!\Gamma(2l+2+k)}\left(\dfrac{r_s}{r_s-r}\right)^k
\end{equation}
Now we need to take the limit $l\rightarrow l-\delta$
\begin{equation}
    \begin{aligned}
        &\Delta_{2l+1+k}=-(-1)^ll!\dfrac{\Gamma(l+1+k)\Gamma(l+1+2ip+k)}{\Gamma(-l+2ip)}\eta\dfrac{l(l+1)}{4ip}.
    \end{aligned}
    \label{Delta1DynDs}
\end{equation}
Since the expression $\Delta_{2l+1}$ is already of first order in $\eta$, we only need $\Delta_{2l+1+k}/\Delta_{2l+1}$ to zeroth order in $\eta$
\begin{equation}
    \Delta_{2l+1+k}/\Delta_{2l+1}=\dfrac{\Gamma(l+1+k)\Gamma(l+1+2ip+k)}{\Gamma(l+1)\Gamma(l+1+2ip)}
    \label{deltadelta}
\end{equation}

Let us now consider the function $Z^{\text{response}}(r)$ \eqref{sourceresponsesplitdynDS}
\begin{equation}
    \begin{aligned}
        &Z^{\text{response}}(r)=\left(1-\dfrac{r_s}{r}\right)^{-l-1}\text{HeunG}\left(\eta,(l+1)^2+\dfrac{l+1}{2}(\eta+4ip)\left|l+1+ip,l+\dfrac{3}{2}+ip,2l+2,\dfrac{1}{2}\right|\dfrac{\eta r_s}{r_s-r}\right)=
        \\
        &=\left(1-\dfrac{r_s}{r}\right)^{-l-1}\sum_{k=0}^{\infty}\dfrac{\Delta'_k\Gamma(2l+2)}{k!\Gamma(2l+2+k)}\left(\dfrac{r_s}{r_s-r}\right)^k,
    \end{aligned}
\end{equation}
where $\Delta'_k$ satisfies the recurrence relation
\begin{equation}
    \Delta'_k=\left(q'+Q'_{k-1}\right)\Delta'_{k-1}-P'_{k-1}R'_{k-2}\Delta'_{k-2},\quad \Delta'_0=1,\quad\Delta'_{-1}=0.
    \label{recnonrot1}
\end{equation}
For our set of parameters, the coefficients take the form:
\begin{gather}
    R'_{k}=\eta(k+1)(k+2l+2)\nonumber\\
    q'+Q'_k=(k+l+1)(k+l+1+2ip)+\eta\left[\dfrac{l+1}{2}+k\left(k+2l+2-\dfrac{1}{2}\right)\right]\\
    P'_{k}=(k+l+ip)\left(k+l+\dfrac{1}{2}+ip\right)\nonumber.
\end{gather}

Since $Z^{\text{response}}(r)$ and $Z^{\text{source}}(r)$ are mapped into one another under $l\leftrightarrow -l-1$, we can immediately write down the solution for $\Delta'_k$ to zeroth order in $\eta$
\begin{equation}
    \Delta'_k=\dfrac{\Gamma(l+1+k)\Gamma(l+1+2ip+k)}{\Gamma(l+1)\Gamma(l+1+2ip)}+\mathcal{O}(\eta)=\dfrac{\Delta_{2l+1+k}}{\Delta_{2l+1}}+\mathcal{O}(\eta).
\end{equation}

Finally, we obtain 
\begin{equation}
    Z^{\text{source}}(r)=\dfrac{1}{\delta}\eta\dfrac{(l!)^2l(l+1)}{4(2l)!(2l+1)!}\prod_{k=1}^{l}\left(k^2+4p^2\right)\left(\dfrac{r_s}{r}\right)^{2l+1}Z^{\text{response}}(r)+\mathcal{O}(\delta^0).
\end{equation}

\section{Wave basis}
To compute the Love numbers for the sub-throat modes, we need to introduce a basis of functions that describes waves propagating in both directions near the throat. 

\subsection{Thin-shell wormhole case}\label{app:SCHwavebasis}

The general solution of the radial equation \eqref{SchWHscalareqnnearzone} can be written in terms of hypergeometric functions
\begin{equation}
    \begin{aligned}
        &R(r)=A_1F_{-}(r)+A_2F_{+}(r),\quad\text{where}
        \\
        &F_{-}(r)=\left(\dfrac{r}{r_s}\right)^l\left(1-\dfrac{r_s}{r}\right)^{-i\omega r_s}{}_2F_1\left(-l,-l-2i\omega r_s,1-2i\omega r_s;1-\dfrac{r_s}{r}\right)
        \\
        &F_{+}(r)=\left(\dfrac{r}{r_s}\right)^l\left(1-\dfrac{r_s}{r}\right)^{i\omega r_s}{}_2F_1\left(-l,-l+2i\omega r_s,1+2i\omega r_s;1-\dfrac{r_s}{r}\right),
    \end{aligned}
    \label{FmFp}
\end{equation}
which at first sight appear to describe ingoing $F_{-}(z_{\ast})$ and outgoing $F_{+}(z_{\ast})$ waves in the tortoise coordinate \eqref{SchWHtortoisecoordinate}. However, this interpretation is valid at the horizon ($r=r_s$). In the present work, we have a throat ($r=r_0=r_s(1+\epsilon)$) instead of a horizon, and therefore we need to construct the wave basis at this point. 

To do so, we impose the following near-throat behavior for $\Psi(z_{\ast})=r(z_{\ast})R(z_{\ast})$; see Eq.~\eqref{SchWHeffpot}
\begin{equation}
    \Psi(z_{\ast}\sim 0)\sim C_1e^{-i\frac{\chi}{r_s}z_{\ast}}+C_2e^{i\frac{\chi}{r_s}z_{\ast}}.
\end{equation}
The coefficients $C_1$ and $C_2$ are then determined by the value of the function and its derivative at the throat
\begin{equation}
    \begin{aligned}
        &C_1=\dfrac{1}{2}\left(\Psi(0)+i\dfrac{r_s}{\chi}\dfrac{\text{d}}{\text{d}z_{\ast}}\Psi(0)\right)=\dfrac{1}{2}\left(r_0R(r_0)+i\dfrac{r_s}{\chi}\dfrac{\epsilon}{1+\epsilon}\left[R(r_0)+r_0\dfrac{\text{d}}{\text{d}r}R(r_0)\right]\right) 
        \\
        &C_2=\dfrac{1}{2}\left(\Psi(0)-i\dfrac{r_s}{\chi}\dfrac{\text{d}}{\text{d}z_{\ast}}\Psi(0)\right)=\dfrac{1}{2}\left(r_0R(r_0)-i\dfrac{r_s}{\chi}\dfrac{\epsilon}{1+\epsilon}\left[R(r_0)+r_0\dfrac{\text{d}}{\text{d}r}R(r_0)\right]\right),
    \end{aligned}
    \label{C1C2}
\end{equation}
where we have used the definition of the tortoise coordinate \eqref{SchWHtortoisecoordinate} in the right universe
\begin{equation}
    \dfrac{\text{d}z_{\ast}}{\text{d}r}=\frac{r}{r-r_s}.
\end{equation}

Using \eqref{FmFp} and \eqref{C1C2}, we obtain the following expressions for the basis functions describing ingoing and outgoing waves at the throat
\begin{equation}
    \begin{aligned}
        &W_{\text{in}}(r)=F_{-}(r)-\left(\dfrac{\epsilon}{1+\epsilon}\right)^{-2i\omega r_s}\dfrac{(1+\epsilon)\left(\chi(1+\epsilon)^2-\omega r_s-i(l+1)\epsilon\right)H_{-}\left(\frac{\epsilon}{1+\epsilon}\right)-i\epsilon H'_{-}\left(\frac{\epsilon}{1+\epsilon}\right)}{(1+\epsilon)\left(\chi(1+\epsilon)^2+\omega r_s-i(l+1)\epsilon\right)H_{+}\left(\frac{\epsilon}{1+\epsilon}\right)-i\epsilon H'_{+}\left(\frac{\epsilon}{1+\epsilon}\right)}F_{+}(r)
        \\[6pt]
        &W_{\text{out}}(r)=F_{+}(r)-\left(\dfrac{\epsilon}{1+\epsilon}\right)^{2i\omega r_s}\dfrac{(1+\epsilon)\left(\chi(1+\epsilon)^2-\omega r_s+i(l+1)\epsilon\right)H_{+}\left(\frac{\epsilon}{1+\epsilon}\right)+i\epsilon H'_{+}\left(\frac{\epsilon}{1+\epsilon}\right)}{(1+\epsilon)\left(\chi(1+\epsilon)^2+\omega r_s+i(l+1)\epsilon\right)H_{-}\left(\frac{\epsilon}{1+\epsilon}\right)+i\epsilon H'_{-}\left(\frac{\epsilon}{1+\epsilon}\right)}F_{-}(r),
    \end{aligned}
\end{equation}
where $H_{\pm}(x)={}_2F_1\left(-l,-l\pm 2i\omega r_s,1\pm 2i\omega r_s;x\right)$.


\subsection{Damour-Solodukhin wormhole case}\label{app:DSwavebasis}

The general solution of the radial equation \eqref{DSWHeqnearZone} is written in terms of the even/odd basis
\begin{equation}
    R(z) = (1-z)^{ip}\left[A_1 u^{(0)}_{-}(z)+A_2 u^{(0)}_{+}(z)\right].
\end{equation}

We require the solution of \eqref{DSWHeffectivepotential} to describe propagating waves inside the throat, so that near the throat it takes the form
\begin{equation}
    \Psi(z_{\ast}\sim 0)\sim C_1e^{-i\frac{\chi}{r_s}z_{\ast}}+C_2e^{i\frac{\chi}{r_s}z_{\ast}}.
\end{equation}
The coefficients $C_1$ and $C_2$ can then be expressed in terms of the value of the solution $\Psi(z_{\ast}=0)$ and its derivative $\frac{\text{d}}{\text{d}z_{\ast}}\Psi(z_{\ast}=0)$ at the throat as
\begin{equation}
    C_1=\dfrac{1}{2}\left(\Psi(0)+i\dfrac{r_s}{\chi}\dfrac{\text{d}}{\text{d}z_{\ast}}\Psi(0)\right),\quad C_2=\dfrac{1}{2}\left(\Psi(0)-i\dfrac{r_s}{\chi}\dfrac{\text{d}}{\text{d}z_{\ast}}\Psi(0)\right).
\end{equation}
The dimensionless coordinate $z$ and the tortoise coordinate \eqref{DSWHtortoisecoordinate} in the right universe are related by 
\begin{equation}
    \dfrac{\text{d}z}{\text{d}z_{\ast}}=-\dfrac{\sqrt{z(z-\eta)}}{r_s(1-z)}.
\end{equation}
Using this, we obtain the following relation between the wave basis coefficients and even/odd basis coefficients 
\begin{equation}
    C_1=\dfrac{r_s}{2}\left(A_1+\dfrac{\eta^{\frac{1}{2}}}{2\chi}A_2\right),\quad C_2=\dfrac{r_s}{2}\left(A_1-\dfrac{\eta^{\frac{1}{2}}}{2\chi}A_2\right).
\end{equation}

Therefore, the basis functions describing waves propagating in both directions inside the throat of the DS wormhole take the form
\begin{equation}
    W_{\text{in}}(z)=(1-z)^{ip}\left[u^{(0)}_{-}(z)+2\chi\eta^{-\frac{1}{2}}u^{(0)}_{+}(z)\right],\quad W_{\text{out}}(z)=(1-z)^{ip}\left[u^{(0)}_{-}(z)-2\chi\eta^{-\frac{1}{2}}u^{(0)}_{+}(z)\right]
\end{equation}


\newpage
\begin{thebibliography}{999}

\bibitem{LIGOScientific:2016sjg}
B.~P.~Abbott \textit{et al.} [LIGO Scientific and Virgo],
Phys. Rev. Lett. \textbf{116} (2016) no.24, 241103
\href{https://arxiv.org/abs/1606.04855}{[arXiv:1606.04855 [gr-qc]]}.

\bibitem{LIGOScientific:2016aoc}
B.~P.~Abbott \textit{et al.} [LIGO Scientific and Virgo],
Phys. Rev. Lett. \textbf{116} (2016) no.6, 061102
\href{https://arxiv.org/abs/1602.03837}{[arXiv:1602.03837 [gr-qc]]}.

\bibitem{KAGRA:2021vkt}
R.~Abbott \textit{et al.} [KAGRA, VIRGO and LIGO Scientific],
Phys. Rev. X \textbf{13} (2023) no.4, 041039
\href{https://arxiv.org/abs/2111.03606}{[arXiv:2111.03606 [gr-qc]]}.

\bibitem{LIGOScientific:2020tif}
R.~Abbott \textit{et al.} [LIGO Scientific and Virgo],
Phys. Rev. D \textbf{103} (2021) no.12, 122002
\href{https://arxiv.org/abs/2010.14529}{[arXiv:2010.14529 [gr-qc]]}.

\bibitem{Blanchet:2013haa}
L.~Blanchet,
Living Rev. Rel. \textbf{17} (2014), 2
\href{https://arxiv.org/abs/1310.1528}{[arXiv:1310.1528 [gr-qc]]}.

\bibitem{Buonanno:1998gg}
A.~Buonanno and T.~Damour,
Phys. Rev. D \textbf{59} (1999), 084006
\href{https://arxiv.org/abs/gr-qc/9811091}{[arXiv:gr-qc/9811091 [gr-qc]]}.

\bibitem{Blanchet:2009sd}
L.~Blanchet, S.~L.~Detweiler, A.~Le Tiec and B.~F.~Whiting,
Phys. Rev. D \textbf{81} (2010), 064004
\href{https://arxiv.org/abs/0910.0207}{[arXiv:0910.0207 [gr-qc]]}.

\bibitem{Campanelli:2005dd}
M.~Campanelli, C.~O.~Lousto, P.~Marronetti and Y.~Zlochower,
Phys. Rev. Lett. \textbf{96} (2006), 111101
\href{https://arxiv.org/abs/gr-qc/0511048}{[arXiv:gr-qc/0511048 [gr-qc]]}.

\bibitem{Vishveshwara:1970zz}
C.~V.~Vishveshwara,
\href{https://www.nature.com/articles/227936a0}{Nature \textbf{227} (1970), 936-938}.

\bibitem{Nollert:1999ji}
H.~P.~Nollert,
Class. Quant. Grav. \textbf{16} (1999), R159-R216.

\bibitem{Dreyer:2003bv}
O.~Dreyer, B.~J.~Kelly, B.~Krishnan, L.~S.~Finn, D.~Garrison and R.~Lopez-Aleman,
Class. Quant. Grav. \textbf{21} (2004), 787-804
\href{https://arxiv.org/abs/gr-qc/0309007}{[arXiv:gr-qc/0309007 [gr-qc]]}.

\bibitem{Berti:2009kk}
E.~Berti, V.~Cardoso and A.~O.~Starinets,
Class. Quant. Grav. \textbf{26} (2009), 163001
\href{https://arxiv.org/abs/0905.2975}{[arXiv:0905.2975 [gr-qc]]}.

\bibitem{Konoplya:2011qq}
R.~A.~Konoplya and A.~Zhidenko,
Rev. Mod. Phys. \textbf{83} (2011), 793-836
\href{https://arxiv.org/abs/1102.4014}{[arXiv:1102.4014 [gr-qc]]}.

\bibitem{Flanagan:2007ix}
E.~E.~Flanagan and T.~Hinderer,
Phys. Rev. D \textbf{77} (2008), 021502
\href{https://arxiv.org/abs/0709.1915}{[arXiv:0709.1915 [astro-ph]]}.

\bibitem{Hinderer:2007mb}
T.~Hinderer,
Astrophys. J. \textbf{677} (2008), 1216-1220
[erratum: Astrophys. J. \textbf{697} (2009) no.1, 964]
\href{https://arxiv.org/abs/0711.2420}{[arXiv:0711.2420 [astro-ph]]}.

\bibitem{Hinderer:2009ca}
T.~Hinderer, B.~D.~Lackey, R.~N.~Lang and J.~S.~Read,
Phys. Rev. D \textbf{81}, 123016 (2010),
\href{https://arxiv.org/abs/0911.3535}{[arXiv:0911.3535 [astro-ph.HE]]}.

\bibitem{Fang:2005qq}
H.~Fang and G.~Lovelace,
Phys. Rev. D \textbf{72} (2005), 124016
\href{https://arxiv.org/abs/gr-qc/0505156}{[arXiv:gr-qc/0505156 [gr-qc]]}.

\bibitem{Binnington:2009bb}
T.~Binnington and E.~Poisson,
Phys. Rev. D \textbf{80} (2009), 084018
\href{https://arxiv.org/abs/0906.1366}{[arXiv:0906.1366 [gr-qc]]}.

\bibitem{Damour:2009vw}
T.~Damour and A.~Nagar,
Phys. Rev. D \textbf{80} (2009), 084035
\href{https://arxiv.org/abs/0906.0096}{[arXiv:0906.0096 [gr-qc]]}.

\bibitem{Kol:2011vg}
B.~Kol and M.~Smolkin,
JHEP \textbf{02} (2012), 010
\href{https://arxiv.org/abs/1110.3764}{[arXiv:1110.3764 [hep-th]]}.

\bibitem{Hui:2020xxx}
L.~Hui, A.~Joyce, R.~Penco, L.~Santoni and A.~R.~Solomon,
JCAP \textbf{04} (2021), 052
\href{https://arxiv.org/abs/2010.00593}{[arXiv:2010.00593 [hep-th]]}.

\bibitem{Poisson:2014gka}
E.~Poisson,
Phys. Rev. D \textbf{91} (2015) no.4, 044004
\href{https://arxiv.org/abs/1411.4711}{[arXiv:1411.4711 [gr-qc]]}.

\bibitem{LeTiec:2020spy}
A.~Le Tiec and M.~Casals,
Phys. Rev. Lett. \textbf{126} (2021) no.13, 131102
\href{https://arxiv.org/abs/2007.00214}{[arXiv:2007.00214 [gr-qc]]}.

\bibitem{LeTiec:2020bos}
A.~Le Tiec, M.~Casals and E.~Franzin,
Phys. Rev. D \textbf{103} (2021) no.8, 084021
\href{https://arxiv.org/abs/2010.15795}{[arXiv:2010.15795 [gr-qc]]}.

\bibitem{Charalambous:2021mea}
P.~Charalambous, S.~Dubovsky and M.~M.~Ivanov,
JHEP \textbf{05} (2021), 038
\href{https://arxiv.org/abs/2102.08917}{[arXiv:2102.08917 [hep-th]]}.

\bibitem{Charalambous:2021kcz}
P.~Charalambous, S.~Dubovsky and M.~M.~Ivanov,
Phys. Rev. Lett. \textbf{127} (2021) no.10, 101101
\href{https://arxiv.org/abs/2103.01234}{[arXiv:2103.01234 [hep-th]]}.

\bibitem{Charalambous:2022rre}
P.~Charalambous, S.~Dubovsky and M.~M.~Ivanov,
JHEP \textbf{10} (2022), 175
\href{https://arxiv.org/abs/2209.02091}{[arXiv:2209.02091 [hep-th]]}.

\bibitem{Sharma:2024hlz}
C.~Sharma, R.~Ghosh and S.~Sarkar,
Phys. Rev. D \textbf{109} (2024) no.4, 4
\href{https://arxiv.org/abs/2401.00703}{[arXiv:2401.00703 [gr-qc]]}.

\bibitem{Ghosh:2026vig}
R.~Ghosh, R.~P.~Bhatt, S.~Chakraborty and S.~Bose,
\href{https://arxiv.org/abs/2604.06249}{[arXiv:2604.06249 [gr-qc]]}.

\bibitem{Cardoso:2017cfl}
V.~Cardoso, E.~Franzin, A.~Maselli, P.~Pani and G.~Raposo,
Phys. Rev. D \textbf{95} (2017) no.8, 084014
\href{https://arxiv.org/abs/1701.01116}{[arXiv:1701.01116 [gr-qc]]}.

\bibitem{Mendes:2016vdr}
R.~F.~P.~Mendes and H.~Yang,
Class. Quant. Grav. \textbf{34} (2017) no.18, 185001
\href{https://arxiv.org/abs/1606.03035}{[arXiv:1606.03035 [astro-ph.CO]]}.

\bibitem{Pani:2015tga}
P.~Pani,
Phys. Rev. D \textbf{92} (2015) no.12, 124030
[erratum: Phys. Rev. D \textbf{95} (2017) no.4, 049902]
\href{https://arxiv.org/abs/1506.06050}{[arXiv:1506.06050 [gr-qc]]}.

\bibitem{Chakraborty:2023zed}
S.~Chakraborty, E.~Maggio, M.~Silvestrini and P.~Pani,
Phys. Rev. D \textbf{110} (2024) no.8, 084042
\href{https://arxiv.org/abs/2310.06023}{[arXiv:2310.06023 [gr-qc]]}.

\bibitem{Cardoso:2018ptl}
V.~Cardoso, M.~Kimura, A.~Maselli and L.~Senatore,
Phys. Rev. Lett. \textbf{121} (2018) no.25, 251105
[erratum: Phys. Rev. Lett. \textbf{131} (2023) no.10, 109903]
\href{https://arxiv.org/abs/1808.08962}{[arXiv:1808.08962 [gr-qc]]}.

\bibitem{Barbosa:2025uau}
S.~Barbosa, P.~Brax, S.~Fichet and L.~de Souza,
JCAP \textbf{07} (2025), 071
\href{https://arxiv.org/abs/2501.18684}{[arXiv:2501.18684 [hep-th]]}.

\bibitem{Cano:2025zyk}
P.~A.~Cano,
JHEP \textbf{07} (2025), 152
\href{https://arxiv.org/abs/2502.20185}{[arXiv:2502.20185 [gr-qc]]}.

\bibitem{Chakravarti:2018vlt}
K.~Chakravarti, S.~Chakraborty, S.~Bose and S.~SenGupta,
Phys. Rev. D \textbf{99} (2019) no.2, 024036
\href{https://arxiv.org/abs/1811.11364}{[arXiv:1811.11364 [gr-qc]]}.

\bibitem{Pereniguez:2021xcj}
D.~Pere{\~n}iguez and V.~Cardoso,
Phys. Rev. D \textbf{105} (2022) no.4, 044026
\href{https://arxiv.org/abs/2112.08400}{[arXiv:2112.08400 [gr-qc]]}.

\bibitem{Charalambous:2023jgq}
P.~Charalambous and M.~M.~Ivanov,
JHEP \textbf{07} (2023), 222
\href{https://arxiv.org/abs/2303.16036}{[arXiv:2303.16036 [hep-th]]}.

\bibitem{Rodriguez:2023xjd}
M.~J.~Rodriguez, L.~Santoni, A.~R.~Solomon and L.~F.~Temoche,
Phys. Rev. D \textbf{108} (2023) no.8, 8
\href{https://arxiv.org/abs/2304.03743}{[arXiv:2304.03743 [hep-th]]}.

\bibitem{Cardoso:2019upw}
V.~Cardoso and F.~Duque,
Phys. Rev. D \textbf{101} (2020) no.6, 064028
\href{https://arxiv.org/abs/1912.07616}{[arXiv:1912.07616 [gr-qc]]}.

\bibitem{Cannizzaro:2024fpz}
E.~Cannizzaro, V.~De Luca and P.~Pani,
Phys. Rev. D \textbf{110} (2024) no.12, 123004
\href{https://arxiv.org/abs/2408.14208}{[arXiv:2408.14208 [astro-ph.HE]]}.

\bibitem{Barbosa:2026qcv}
S.~Barbosa, S.~Fichet and L.~de Souza,
JCAP \textbf{09} (2026), 009
\href{https://arxiv.org/abs/2602.00349}{[arXiv:2602.00349 [hep-th]]}.

\bibitem{Love:1909}
A.~E.~H.~Love,
\href{https://royalsocietypublishing.org/rspa/article/82/551/73/4071/The-yielding-of-the-earth-to-disturbing-forces}{Proc. Roy. Soc. Lond. A \textbf{82}, 73--88 (1909)}.

\bibitem{Cardoso:2008bp}
V.~Cardoso, A.~S.~Miranda, E.~Berti, H.~Witek and V.~T.~Zanchin,
Phys. Rev. D \textbf{79} (2009) no.6, 064016
\href{https://arxiv.org/abs/0812.1806}{[arXiv:0812.1806 [hep-th]]}.

\bibitem{Konoplya:2017wot}
R.~A.~Konoplya and Z.~Stuchl{\'\i}k,
Phys. Lett. B \textbf{771} (2017), 597-602
\href{https://arxiv.org/abs/1705.05928}{[arXiv:1705.05928 [gr-qc]]}.

\bibitem{Abramowicz:2002vt}
M.~A.~Abramowicz, W.~Kluzniak and J.~P.~Lasota,
Astron. Astrophys. \textbf{396} (2002), L31-L34
\href{https://arxiv.org/abs/astro-ph/0207270}{[arXiv:astro-ph/0207270 [astro-ph]]}.

\bibitem{Mazur:2001fv}
P.~O.~Mazur and E.~Mottola,
Universe \textbf{9} (2023) no.2, 88
\href{https://arxiv.org/abs/gr-qc/0109035}{[arXiv:gr-qc/0109035 [gr-qc]]}.

\bibitem{Schunck:2003kk}
F.~E.~Schunck and E.~W.~Mielke,
Class. Quant. Grav. \textbf{20} (2003), R301-R356
\href{https://arxiv.org/abs/0801.0307}{[arXiv:0801.0307 [astro-ph]]}.

\bibitem{Herdeiro:2021lwl}
C.~A.~R.~Herdeiro, A.~M.~Pombo, E.~Radu, P.~V.~P.~Cunha and N.~Sanchis-Gual,
JCAP \textbf{04} (2021), 051
\href{https://arxiv.org/abs/2102.01703}{[arXiv:2102.01703 [gr-qc]]}.

\bibitem{Abramowicz:1997qk}
M.~A.~Abramowicz, M.~Bruni, S.~Sonego, N.~Andersson and P.~Ghosh,
\href{https://iopscience.iop.org/article/10.1088/0264-9381/14/12/002}{Class. Quant. Grav. \textbf{14} (1997), L189-L194}.

\bibitem{Morris:1988tu}
M.~S.~Morris, K.~S.~Thorne and U.~Yurtsever,
\href{https://journals.aps.org/prl/abstract/10.1103/PhysRevLett.61.1446}{Phys. Rev. Lett. \textbf{61} (1988), 1446-1449}.

\bibitem{Lemos:2008cv}
J.~P.~S.~Lemos and O.~B.~Zaslavskii,
Phys. Rev. D \textbf{78} (2008), 024040
\href{https://arxiv.org/abs/0806.0845}{[arXiv:0806.0845 [gr-qc]]}.

\bibitem{Damour:2007ap}
T.~Damour and S.~N.~Solodukhin,
Phys. Rev. D \textbf{76} (2007), 024016
\href{https://arxiv.org/abs/0704.2667}{[arXiv:0704.2667 [gr-qc]]}.

\bibitem{Mathur:2005zp}
S.~D.~Mathur,
Fortsch. Phys. \textbf{53} (2005), 793-827
\href{https://arxiv.org/abs/hep-th/0502050}{[arXiv:hep-th/0502050 [hep-th]]}.

\bibitem{Cardoso:2019rvt} V.~Cardoso and P.~Pani, 
Living Rev. Rel. \textbf{22} (2019) no.1, 4 
\href{https://arxiv.org/abs/1904.05363}{[arXiv:1904.05363 [gr-qc]]}.

\bibitem{Visser:1989kg}
M.~Visser,
Nucl. Phys. B \textbf{328} (1989), 203-212
\href{https://arxiv.org/abs/0809.0927}{[arXiv:0809.0927 [gr-qc]]}.

\bibitem{Visser:1995cc}
M.~Visser,
``Lorentzian wormholes: From Einstein to Hawking,''
AIP Press, Woodbury, New York (1995), 412 pp.

\bibitem{Harko:2013yb}
T.~Harko, F.~S.~N.~Lobo, M.~K.~Mak and S.~V.~Sushkov,
Phys. Rev. D \textbf{87} (2013) no.6, 067504
\href{https://arxiv.org/abs/1301.6878}{[arXiv:1301.6878 [gr-qc]]}.

\bibitem{Moraes:2017dbs}
P.~H.~R.~S.~Moraes and P.~K.~Sahoo,
Phys. Rev. D \textbf{97} (2018) no.2, 024007
\href{https://arxiv.org/abs/1709.00027}{[arXiv:1709.00027 [gr-qc]]}.

\bibitem{Svitek:2016nvm}
O.~Svitek and T.~Tahamtan,
Eur. Phys. J. C \textbf{78} (2018) no.2, 167
\href{https://arxiv.org/abs/1606.01501}{[arXiv:1606.01501 [gr-qc]]}.

\bibitem{Garattini:2019ivd}
R.~Garattini,
Eur. Phys. J. C \textbf{79} (2019) no.11, 951
\href{https://arxiv.org/abs/1907.03623}{[arXiv:1907.03623 [gr-qc]]}.

\bibitem{Berthiere:2017tms}
C.~Berthiere, D.~Sarkar and S.~N.~Solodukhin,
Phys. Lett. B \textbf{786} (2018), 21-27
\href{https://arxiv.org/abs/1712.09914}{[arXiv:1712.09914 [hep-th]]}.

\bibitem{Potaux:2021yan}
Y.~Potaux, D.~Sarkar and S.~N.~Solodukhin,
Phys. Rev. D \textbf{105} (2022) no.2, 025015
\href{https://arxiv.org/abs/2112.03855}{[arXiv:2112.03855 [hep-th]]}.

\bibitem{Potaux:2022uxa}
Y.~Potaux, D.~Sarkar and S.~N.~Solodukhin,
Phys. Rev. Lett. \textbf{130} (2023) no.26, 261501
\href{https://arxiv.org/abs/2212.13208}{[arXiv:2212.13208 [hep-th]]}.

\bibitem{Potaux:2023fwm}
Y.~Potaux, D.~Sarkar and S.~N.~Solodukhin,
Phys. Rev. D \textbf{108} (2023) no.12, 125012
\href{https://arxiv.org/abs/2310.18745}{[arXiv:2310.18745 [hep-th]]}.

\bibitem{Wielgus:2020uqz}
M.~Wielgus, J.~Horak, F.~Vincent and M.~Abramowicz,
Phys. Rev. D \textbf{102} (2020) no.8, 084044
\href{https://arxiv.org/abs/2008.10130}{[arXiv:2008.10130 [gr-qc]]}.

\bibitem{Wang:2020emr}
X.~Wang, P.~C.~Li, C.~Y.~Zhang and M.~Guo,
Phys. Lett. B \textbf{811} (2020), 135930
\href{https://arxiv.org/abs/2007.03327}{[arXiv:2007.03327 [gr-qc]]}.

\bibitem{Guerrero:2022qkh} 
M.~Guerrero, G.~J.~Olmo, D.~Rubiera-Garcia and D.~G{\'o}mez S{\'a}ez-Chill{\'o}n, 
Phys. Rev. D \textbf{105} (2022) no.8, 084057
\href{https://arxiv.org/abs/2202.03809}{[arXiv:2202.03809 [gr-qc]]}.

\bibitem{Guerrero:2021pxt}
M.~Guerrero, G.~J.~Olmo and D.~Rubiera-Garcia,
JCAP \textbf{04} (2021), 066
\href{https://arxiv.org/abs/2102.00840}{[arXiv:2102.00840 [gr-qc]]}.

\bibitem{Solodukhin:2025opw}
S.~N.~Solodukhin and V.~Tagiev,
Phys. Rev. D \textbf{113} (2026) no.6, 064017
\href{https://arxiv.org/abs/2511.03879}{[arXiv:2511.03879 [gr-qc]]}.

\bibitem{Cardoso:2016oxy} V.~Cardoso, S.~Hopper, C.~F.~B.~Macedo, C.~Palenzuela and P.~Pani, 
Phys. Rev. D \textbf{94} (2016) no.8, 084031
\href{https://arxiv.org/abs/1608.08637}{[arXiv:1608.08637 [gr-qc]]}.

\bibitem{Cardoso:2016rao}
V.~Cardoso, E.~Franzin and P.~Pani,
Phys. Rev. Lett. \textbf{116} (2016) no.17, 171101
[erratum: Phys. Rev. Lett. \textbf{117} (2016) no.8, 089902]
\href{https://arxiv.org/abs/1602.07309}{[arXiv:1602.07309 [gr-qc]]}.

\bibitem{Hui:2019aox}
L.~Hui, D.~Kabat and S.~S.~C.~Wong,
JCAP \textbf{12} (2019), 020
\href{https://arxiv.org/abs/1909.10382}{[arXiv:1909.10382 [gr-qc]]}.

\bibitem{Abedi:2016hgu}
J.~Abedi, H.~Dykaar and N.~Afshordi,
Phys. Rev. D \textbf{96} (2017) no.8, 082004
\href{https://arxiv.org/abs/1612.00266}{[arXiv:1612.00266 [gr-qc]]}.

\bibitem{Bueno:2017hyj}
P.~Bueno, P.~A.~Cano, F.~Goelen, T.~Hertog and B.~Vercnocke,
Phys. Rev. D \textbf{97} (2018) no.2, 024040
\href{https://arxiv.org/abs/1711.00391}{[arXiv:1711.00391 [gr-qc]]}.

\bibitem{Chakraborty:2026qru}
S.~Chakraborty and P.~Pani,
\href{https://arxiv.org/abs/2604.08679}{[arXiv:2604.08679 [gr-qc]]}.


\bibitem{Uchikata:2016qku}
N.~Uchikata, S.~Yoshida and P.~Pani,
Phys. Rev. D \textbf{94} (2016) no.6, 064015
\href{https://arxiv.org/abs/1607.03593}{[arXiv:1607.03593 [gr-qc]]}.

\bibitem{Biswas:2026vdp}
S.~Biswas,
Phys. Rev. D \textbf{114} (2026) no.6, 064050
\href{https://arxiv.org/abs/2605.03025}{[arXiv:2605.03025 [gr-qc]]}.

\bibitem{Giri:2024cks}
S.~Giri, U.~Danielsson, L.~Lehner and F.~Pretorius,
Phys. Rev. D \textbf{111} (2025) no.2, 024007
\href{https://arxiv.org/abs/2405.08062}{[arXiv:2405.08062 [gr-qc]]}.

\bibitem{Nair:2022xfm}
S.~Nair, S.~Chakraborty and S.~Sarkar,
Phys. Rev. D \textbf{107} (2023) no.12, 124041
\href{https://arxiv.org/abs/2208.06235}{[arXiv:2208.06235 [gr-qc]]}.

\bibitem{Chakrabarti:2013xza}
S.~Chakrabarti, T.~Delsate and J.~Steinhoff,
Phys. Rev. D \textbf{88} (2013), 084038
\href{https://arxiv.org/abs/1306.5820}{[arXiv:1306.5820 [gr-qc]]}.

\bibitem{Yunes:2005ve}
N.~Yunes and J.~Gonzalez,
Phys. Rev. D \textbf{73} (2006) no.2, 024010
[erratum: Phys. Rev. D \textbf{89} (2014) no.8, 089902]
\href{https://arxiv.org/abs/gr-qc/0510076}{[arXiv:gr-qc/0510076 [gr-qc]]}.

\bibitem{Gralla:2017djj}
S.~E.~Gralla,
Class. Quant. Grav. \textbf{35} (2018) no.8, 085002
\href{https://arxiv.org/abs/1710.11096}{[arXiv:1710.11096 [gr-qc]]}.

\bibitem{Goldberger:2004jt}
W.~D.~Goldberger and I.~Z.~Rothstein,
Phys. Rev. D \textbf{73} (2006), 104029
\href{https://arxiv.org/abs/hep-th/0409156}{[arXiv:hep-th/0409156 [hep-th]]}.

\bibitem{Goldberger:2005cd}
W.~D.~Goldberger and I.~Z.~Rothstein,
Phys. Rev. D \textbf{73} (2006), 104030
\href{https://arxiv.org/abs/hep-th/0511133}{[arXiv:hep-th/0511133 [hep-th]]}.

\bibitem{Rodriguez:2026iot}
M.~J.~Rodr{\'\i}guez, L.~Santoni and A.~R.~Solomon,
\href{https://arxiv.org/abs/2604.08653}{[arXiv:2604.08653 [gr-qc]]}.

\bibitem{Ivanov:2022hlo}
M.~M.~Ivanov and Z.~Zhou,
Phys. Rev. D \textbf{107} (2023) no.8, 084030
\href{https://arxiv.org/abs/2208.08459}{[arXiv:2208.08459 [hep-th]]}.

\bibitem{Unruh:1976fm}
W.~G.~Unruh,
``Absorption Cross Section of Small Black Holes,''
\href{https://journals.aps.org/prd/abstract/10.1103/PhysRevD.14.3251}{Phys. Rev. D \textbf{14}, 3251 (1976)}.

\bibitem{Das:1996we}
S.~R.~Das, G.~W.~Gibbons and S.~D.~Mathur,
``Universality of Low Energy Absorption Cross Sections for Black Holes,''
Phys. Rev. Lett. \textbf{78}, 417--419 (1997).
\href{https://arxiv.org/abs/2604.08653}{[arXiv:hep-th/9609052 [hep-th]]}.

\bibitem{Starobinsky:1973aij}
A.~A.~Starobinsky,
``Amplification of waves during reflection from a rotating black hole,''
Sov. Phys. JETP \textbf{37}, 28--32 (1973)
[Zh. Eksp. Teor. Fiz. \textbf{64}, 48--57 (1973)].

\bibitem{Starobinsky:1973aij2}
A.~A.~Starobinsky and S.~M.~Churilov,
``Amplification of electromagnetic and gravitational waves scattered by a rotating black hole,''
Sov. Phys. JETP \textbf{38}, 1--5 (1974)
[Zh. Eksp. Teor. Fiz. \textbf{65}, 3--11 (1973)].

\bibitem{Freivogel:2026ujn}
B.~Freivogel, A.~Fumagalli and M.~Toma{\v{s}}evi{\'c},
\href{https://arxiv.org/abs/2606.12528}{[arXiv:2606.12528 [hep-th]]}.

\bibitem{Combaluzier--Szteinsznaider:2025eoc}
O.~Combaluzier--Szteinsznaider, D.~Glazer, A.~Joyce, M.~J.~Rodriguez and L.~Santoni,
JHEP \textbf{06} (2026), 032
\href{https://arxiv.org/abs/2511.02372}{[arXiv:2511.02372 [gr-qc]]}.

\bibitem{Kobayashi:2025vgl}
H.~Kobayashi, S.~Mukohyama, N.~Oshita, K.~Takahashi and V.~Yingcharoenrat,
Phys. Rev. D \textbf{113} (2026) no.8, 8
\href{https://arxiv.org/abs/2511.12580}{[arXiv:2511.12580 [gr-qc]]}.

\bibitem{Mano:1996vt}
S.~Mano, H.~Suzuki and E.~Takasugi,
Prog. Theor. Phys. \textbf{95} (1996), 1079-1096
\href{https://arxiv.org/abs/gr-qc/9603020}{[arXiv:gr-qc/9603020 [gr-qc]]}.

\bibitem{Mano:1996mf}
S.~Mano, H.~Suzuki and E.~Takasugi,
Prog. Theor. Phys. \textbf{96} (1996), 549-566
\href{https://arxiv.org/abs/gr-qc/9605057}{[arXiv:gr-qc/9605057 [gr-qc]]}.

\bibitem{Fiziev:2005ki}
P.~P.~Fiziev,
Class. Quant. Grav. \textbf{23} (2006), 2447-2468
\href{https://arxiv.org/abs/gr-qc/0509123}{[arXiv:gr-qc/0509123 [gr-qc]]}.

\bibitem{Bonelli:2021uvf}
G.~Bonelli, C.~Iossa, D.~P.~Lichtig and A.~Tanzini,
Phys. Rev. D \textbf{105} (2022) no.4, 044047
\href{https://arxiv.org/abs/2105.04483}{[arXiv:2105.04483 [hep-th]]}.

\bibitem{Bonelli:2022ten}
G.~Bonelli, C.~Iossa, D.~Panea Lichtig and A.~Tanzini,
Commun. Math. Phys. \textbf{397} (2023) no.2, 635-727
\href{https://arxiv.org/abs/2201.04491}{[arXiv:2201.04491 [hep-th]]}.

\bibitem{Lisovyy:2022flm}
O.~Lisovyy and A.~Naidiuk,
J. Phys. A \textbf{55} (2022) no.43, 434005
\href{https://arxiv.org/abs/2208.01604}{[arXiv:2208.01604 [math-ph]]}.

\bibitem{Hui:2022vbh}
L.~Hui, A.~Joyce, R.~Penco, L.~Santoni and A.~R.~Solomon,
JHEP \textbf{09} (2022), 049
\href{https://arxiv.org/abs/2203.08832}{[arXiv:2203.08832 [hep-th]]}.

\bibitem{Israel:1966rt}
W.~Israel,
Nuovo Cim. B \textbf{44S10} (1966), 1
[erratum: Nuovo Cim. B \textbf{48} (1967), 463].

\bibitem{Ivanov:2024sds}
M.~M.~Ivanov, Y.~Z.~Li, J.~Parra-Martinez and Z.~Zhou,
Phys. Rev. Lett. \textbf{132} (2024) no.13, 131401
[erratum: Phys. Rev. Lett. \textbf{134} (2025) no.15, 159901]
\href{https://arxiv.org/abs/2401.08752}{[arXiv:2401.08752 [hep-th]]}.

\bibitem{Poisson:1995sv}
E.~Poisson and M.~Visser,
Phys. Rev. D \textbf{52} (1995), 7318-7321
\href{https://arxiv.org/abs/gr-qc/9506083}{[arXiv:gr-qc/9506083 [gr-qc]]}.

\bibitem{Tsukamoto:2019ihj}
N.~Tsukamoto and T.~Kokubu,
Phys. Rev. D \textbf{101} (2020) no.4, 044030
\href{https://arxiv.org/abs/1912.07492}{[arXiv:1912.07492 [gr-qc]]}.

\end{thebibliography}
\end{document}